\documentclass[10pt,journal,a4paper]{IEEEtran}

\usepackage{cite}
\usepackage{xcolor}
\usepackage{graphicx}
\usepackage[latin1]{inputenc}
\usepackage[T1]{fontenc}
\usepackage{amsmath,amsfonts,amsbsy,amssymb}
\usepackage{mathabx}
\usepackage{url}
\usepackage{commath}

\usepackage{mathrsfs}
\usepackage[nolist]{acronym}
\usepackage{tabularx}
\usepackage{amssymb}
\usepackage{amsmath}
\usepackage{graphicx}
\usepackage[latin1]{inputenc}
\usepackage{graphicx}
\usepackage{cite}
\usepackage{multirow}
\usepackage{wasysym}
\usepackage{multirow}
\usepackage{float}
\usepackage{color}
\usepackage{enumitem}
\usepackage{amsthm}	
\usepackage{xcolor,soul}
\definecolor{aqua}{rgb}{0.0, 1.0, 1.0}
\definecolor{laranja}{rgb}{1.0, 0.6, 0.2}

\newtheorem{proposition}{Proposition}
\newtheorem{lemma}{Lemma}

\begin{document}

\title{A Throughput and Energy Efficiency Scheme for Unlicensed Massive Machine Type Communications}
\author{
	\IEEEauthorblockN{Iran Ramezanipour, Hirley Alves, Pedro J. H. Nardelli and Ari Pouttu} 

	\thanks{I. Ramezanipour, H. Alves and A. Pouttu are with the Centre for Wireless Communications (CWC), University of Oulu, Finland. Contact: firstname.lastname@oulu.fi.}
	\thanks{P. Nardelli is with Lappeenranta University of Technology, Lappeenranta, Finland. Contact: pedro.nardelli@lut.fi.}
	\thanks{This work is partially supported by Aka Project SAFE (Grant n.303532), Strategic Research Council/Aka BCDC Energy (Grant n.$292854$), Finnish Funding Agency for Technology and Innovation (Tekes), Bittium Wireless, Keysight Technologies Finland, Kyynel, MediaTek Wireless, Nokia Solutions and Networks and the Academy of Finland 6Genesis Flagship (Grant no. 318927) and EE-IOT (grant no. 319008).}
}
%
%
\maketitle

\begin{abstract}
In this paper the throughput and energy efficiency of an unlicensed machine type communications network is studied. If an outage event happens in the network, there is a possibility for packet retransmissions in order to obtain a lower error probability. The model consist of a network with two types of users, Licensed and unlicensed users. The licensed users allocated uplink channel is also used by the unlicensed users. However, it is done in a way that no harm is done to the licensed users' transmission from sharing the same channel with the unlicensed users. However, licensed users' transmission causes interference on the unlicensed network. Poisson point process is used here to model the location of the nodes and the effect of interference on the network. We study how different factors such as number of retransmissions, SIR threshold and outage can effect the throughput and energy efficiency of the network. Throughput and energy efficiency are also both studied in constrained optimization problems where the constraints are the SIR threshold and number of retransmission attempts. We also show why it is important to use limited transmissions and what are the benefits.

\end{abstract}


\section{Introduction}

Internet of things has revolutionized the way connection works and is slowly becoming a part of our daily lives. Many applications are currently becoming IoT based, from remotely controlling your house to different processes in an industrial setting \cite{cerwall2015ericsson}. Billions of devices are expected to join the Internet by the year $2020$ which while providing a big economic impact, will also create new challenges such as the availability of spectrum resources \cite{manyika2015unlocking,7004894}. This makes finding ways to efficiently use the spectrum more valuable than ever. Machine-type communications (MTC) is a non-human centric concept introduced under the umbrella of IoT for the future communication technology, $5$G, which can support a high number of connectivity in the network and can provide different quality of services \cite{ali2015next}. 

MTC can be divided into three categories based on the expected properties, (i) enhanced mobile broadband (eMBB) which should be able to provide connectivity with high peak rates in addition to moderate rates for the cell-edge users, (ii) ultra-reliable MTC (uMTC) which focuses on making ultra reliable and low latency connections in the networks possible and (iii) massive machine type communication (mMTC) main goal is to provide massive connectivity for a large number of nodes (in the order of $10$ times higher than the current number of connected devices) with different quality of service (QoS) \cite{8476595, dawy2017toward}. A mMTC network usually consists of billions of low-complexity low-power machine-type devices as nodes. A good example of this type of networks are smart grids where the data from a very large number of nodes (smart meters) needs to be collected\cite{bockelmann2016massive}. Industrial control is also another application of mMTC. In both of these examples, the reliability level of the network needs to be high also since it should be able to handle critical situations \cite{durisi2016toward}. In this paper we focus throughput and energy optimization in a mMTC network in the presence of retransmissions.

The spectrum resources are limited, hence, the availability of spectrum is a never ending challenge for wireless communications. Considering that mMTC is going to connect billions of devices together, this notion is becoming even more challenging in the upcoming $5$G networks, thus, studying different ways to efficiently use the available spectrum is very important. Cognitive radio can provide useful tools which can help the network to use the spectrum more efficiently \cite{akyildiz2006next}. One of these methods is the unlicensed spectrum access which is a suitable option for low-power IoT-based networks and is also the spectrum access method used in this paper. 

Authors in \cite{nardelli2016maximizing, tome2016joint} study the same unlicensed spectrum access model where the unlicensed nodes use the licensed nodes uplink channel too. In there, they show that the position of the unlicensed nodes are fixed which makes it reasonable to use highly directional antennas and limited transmit power in the unlicensed network in order to avoid interference from this network on the licensed network. However, this does not prevent the licensed users causing interference on the unlicensed nodes. Since these works are limited to smart grids applications, we later on expanded these works in \cite{ramezanipour2018increasing}, by following the work done in \cite{nardelli2012optimal, nardelli2014throughput}, to make the model more generalized and compatible with other wireless networks. In this paper, we follow the same model as in \cite{ramezanipour2018increasing, iran, nardelli2016maximizing} to prove that the approximation used there for the average number of retransmission attempt is in fact a tight approximation.  We show why it is worth to use limited number of retransmission, and optimize the throughput as a function of the SIR threshold and number of retransmissions.

Moreover, we also study the energy efficiency of the proposed model in this paper. Energy efficiency is one of the most important problems that needs to be considered in wireless networks, specifically in ultra dense networks \cite{osseiran2014scenarios}. As was previously mentioned, mMTC network are designed to support massive connectivity between billions of IoT devices with minimum human interactions \cite{popovski20185g}, most of which are low powered . Most of theses devices are battery supplied, hence, having a limited energy supply (wireless sensor networks for instance). It also happens often that these mMTC networks are deployed in critical or hard to access locations which makes changing the batteries and renewing the energy resources very difficult \cite{akyildiz2002wireless,de2011energy, vardhan2000wireless}. All this show how important it is to maintain the energy resources in mMTC networks, meaning that energy efficiency is an important issue here that needs to be considered.

Valuable works have been done in the field of energy efficiency. In \cite{hasan2011green}, authors study different challenges and metrics with regards to reducing the total power consumption of the network while in \cite{wang2010energy} and \cite{sadek2009energy} maximizing the energy efficiency by optimizing the packet size and constrained by an outage threshold in non-cooperative and cooperative wireless networks is studied respectively. In \cite{zhihui2014eefa}, a new scheduling algorithm based on frame aggregation is proposed in order to achieve energy efficiency in IEEE $802.11$n wireless networks. In terms of the physical properties of the battery equipped machine type devices, two interesting medium access control protocol and power scheduling schemes are proposed in \cite{ma2007battery, jayashree2004battery} in order ro preserve the battery life. Moreover in \cite{kitahara2008data}, Takeshi Kitahara, et al. introduce a data transmission control method based on the well-known electrochemical characteristics of batteries which makes increasing the discharge capacity possible.

While the above mentioned works are interesting and valuable, none of them really addresses the massive connectivity issue and how the energy efficiency needs to be handled in a mMTC network. In \cite{tu2011energy}, the authors investigate access control algorithms for machine type nodes which can reduce the energy consumption in the uplink channel. In this paper, the machine type nodes access to the base station is maximized by the means of grouping and coordinator selection. However, in this paper we evaluate the energy efficiency of an unlicensed mMTC network and optimize the energy efficiency contained by an outage threshold and maximum number of retransmissions. We also show the effect of different network parameters such as network density and the SIR threshold on the behavior of the energy efficiency.

\subsection{Contributions}

The followings are the main contributions of this paper.
\begin{itemize}
	\item We show how tight the approximation used for the average number of retransmission attempts in our other work is and the desired range that the approximation is valid for.
	
	\item The maximum number of allowed retransmissions and the SIR threshold that leads to the maximum link throughput is studied.
	
	\item The optimal throughput in the sense of spectral efficiency and the optimal energy efficiency are also studied. Also, we show why it is important to have limited transmissions in the network and why it is beneficial to use our proposed optimal throughput model.

\end{itemize}



\section{System Model}\label{sect:smodel} 

\begin{table}
	\caption{Summary of the functions and symbols.}
	\centering
	\begin{tabular}{ | l | l | l }
		\hline
		Symbol & Expression  \\ \hline
		$\Gamma\left[\cdot\right]$ & Gamma Function \\ \hline
		$r_i$ & Distance From the Reference Receiver and the ith Interfering Node  \\ \hline
		$g_i$ & Channel Gain \\ \hline
		$P_\mathrm{p}$ & Licensed Users' Transmit Power \\ \hline
		$P_\mathrm{s}$ & Unlicensed Users' Transmit Power \\ \hline
		SIR & Signal to Interference Ratio\\ \hline
		$\beta$ & SIR Threshold \\ \hline
		$\beta^\ast$ & Optimal SIR Threshold \\ \hline
		$\hat{\Phi}$ & Poisson Point Process\\ \hline
		$P_\textrm{out}$ & Outage Probability \\ \hline
		$P_\textrm{suc}$ & Probability of a Successful Transmission \\ \hline
		$\lambda$ & Network Density (nodes/$m^2$) \\ \hline
		$T$ & Link Throughput\\ \hline 
		$T^\ast$ & Optimal Throughput\\ \hline 
		$\epsilon$ & Outage Threshold\\ \hline
		EE & Energy Efficiency\\ \hline
		$\alpha$ & Path Loss Exponent \\ \hline
		$\textup{P}_{T}$ & Total Power Consumption\\ \hline
		$P_{PA}$ & Power Amplifier Consumed Energy\\ \hline
		$P_{T_x}$ & Transmission Power\\ \hline
		$P_{R_x}$ & Reception Power\\ \hline
		$m$ & Number Of Retransmission Attempts\\ \hline
		$1+\bar{m}$ & Average Number of Retransmission attempts \\ \hline
		$1+\bar{m}_{e}$ & Average Number of Retransmission attempts in the extreme cases \\ \hline
		$m^\ast$ & Maximum Allowed Number Of Retransmission Attempts\\ \hline
		$\delta$ & Drain Efficiency\\ \hline

		\hline
	\end{tabular}
	
	\label{tbl:1}
\end{table}

\begin{figure}[!t]
	\centering
	\includegraphics[width=\columnwidth]{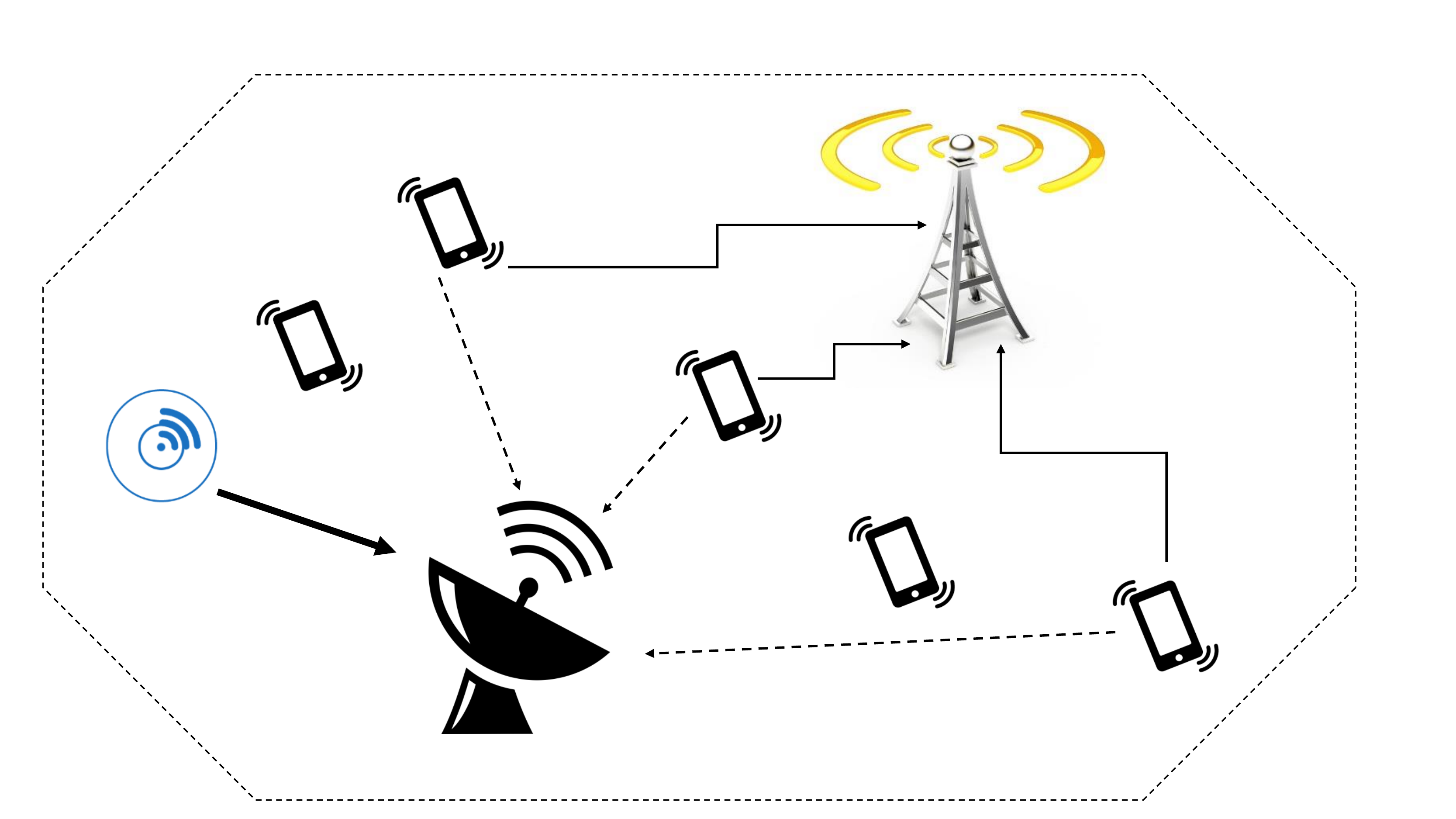}
	\caption{An illustration of the dynamic spectrum access scenario, where licensed and unlicensed users share the up-link channel. The unlicensed transmitter is depicted by the smart meter, the aggregator (unlicensed receiver) by the satellite receiver, the handsets are the mobile licensed users (interferers to the aggregator) and the big antenna is the cellular base-station. As the smart meter uses directional antennas with limited transmit power (bold arrow), its interference towards the base-station can be ignored. The thin black arrows represent the licensed users' desired signal, while the red ones represent their interference towards the aggregator.}
	\label{fig:model}
\end{figure}

In order to do the throughput and energy efficiency analysis in this paper, we use a dynamic spectrum access network model, shown in Fig. \ref{fig:model}. The two types of nodes in this model which are licensed and unlicensed which share the same frequency band which is used by the licensed users in the uplink channel. This sharing is done in a way that the unlicensed users can not cause excess'' any harm or interference to the licensed users. In this model, the licensed link is the one between the mobile users and the cellular base stations while the unlicensed link is what is used by the sensors to communicate with their corresponding aggregators. The unlicensed nodes are considered to be static which would make the use of directional antennas possible, hence, the orientation errors could be avoided \cite{joint}.

Moreover, the power used by the unlicensed users for their transmission is limited. This limitation which is either imposed on them from the licensed network or comes from sensors own properties, restricts the maximum range that the unlicensed network transmitted signal can reach. This means that the unlicensed nodes radiations pattern can be considered as a line segment in which the starting points are these nodes and the end point is determined by the power constraint. Also, it is assumed that in this model, there are no packet collisions between the transmitted packets by the sensors to the same aggregator. This is justifiable since the packet size in these transmissions are assumed to be small and the unlicensed network can take advantage of the multiple access solutions.

Considering the above explanations, we can see that the interference in this model can stem from the following different sources, (i) from the mobile users to aggregators, (ii) from sensors to cellular base stations and (iii) from sensors to aggregators that they are not communicating with. Taking into account the aforementioned system model assumptions, it is possible to conclude that if the licensed and unlicensed nodes are implemented in explicit positions, it makes it possible to eliminate the interferences in cases (i) and (ii). Even if it is assumed that the nodes are implemented in random locations, the probability of having the base station or aggregator in the same line segment as the unlicensed nodes signal is close to zero \cite{ramezanipour2018increasing, iran, nardelli2016maximizing}. As follows, the only source of interference in this model is then (i). In order to be able to evaluate the impact of this interference on the system performance, we need to be able to have a good understanding  of the uncertainty of the licensed node's locations. Hence, a Poisson point process $\Phi$ is used to model the interfering nodes distributed over an infinite two-dimensional plane where $\lambda > 0$ (average number of nodes per $m^2$) denotes the spatial density. This model is also elaborated in details in \cite{ramezanipour2018increasing, haenggi2012stochastic}.

For modeling the wireless channel in this paper, we consider distance dependent path-loss and quasi-static fading model. Here, $r_i$ represents the distance between the reference receiver and the \textit{i}-th interfering node. Based on Slivnyak theorem \cite{haenggi2012stochastic}, the reference receiver being located arbitrarily at the origin means that the receiver's position is at the center of the Euclidean space. Having this location as a fixed point makes locating the surrounding elements easier \cite{ haenggi2012stochastic, iranj}. The channel gain between the reference receiver and the \textit{i}-th interfering node is shown as $g_i$. If we consider $P_{T_x}$ to be the transmit power and $\alpha> 2$, the path-loss exponent, then the power received at the reference receiver is equivalent to $P_{T_x} g_ir_i^{-\alpha}$. With these in hand, the signal to interference ratio at the reference receiver SIR$_0$ is then defined as:

\begin{equation} 
\label{eq_SIR}
\textrm{SIR}_0 = \dfrac{P_\mathrm{s} g_{0}  r_0^{-\alpha}}{P_\mathrm{p} \; \underset{i \in \Phi}{\displaystyle \sum}  g_{i}  r_i^{-\alpha}},
\end{equation}
 
\noindent in which, $P_\mathrm{p}$ denotes the licensed users transmit power and $P_\mathrm{s}$ is the transmit power used by the unlicensed users for their transmissions \footnote{Note that noise is not considered in this analysis, even if it is considered, adding noise would not have much effect on the output as also stated in \cite{weber2010overview}}.
The spectral efficiency in this model is defined as $\log_2(1+\beta)$ in bits/s/Hz. This is justifiable since the reference link takes advantage of point-to-point Gaussian codes and interference-as-noise decoding rules \cite{ramezanipour2018increasing, nardelli2015throughput, baccelli2011interference}. This spectral efficiency can only be achieved if the SIR is greater than a given threshold which here is defined as $\beta$, i.e $\textrm{SIR}>\beta$. An outage event happens if a transmitted message is not successfully decoded at the receiver side, meaning that $\textrm{SIR} \leq \beta$. The probability of the system being in outage is $\mathrm{P}_\textrm{out}=\Pr [\textup{SIR} < \beta]$ . In this model, there is possibility of retransmission in case an outage event occurs. There can be up to $m$ retransmissions in the network, hence, if the message is not successfully decoded by the receiver after $1+m$ transmission attempts (first transmission plus $m$ retransmissions), it is then dropped \cite{p2} which will result in packet loss. Here, $\mathrm{P}_\textrm{suc}$ is used to refer to a probability of having a successfully decoded message and is $\mathrm{P}_\textrm{suc} = 1 - \mathrm{P}_\textrm{out}^{1+m}$.

In order to compute $\mathrm{P}_\textrm{out}$, the previously mentioned channel gains $g$, are considered to be quasi-static (squared envelopes) which are also independent and identically distributed exponentials (Rayleigh fading) with mean $1$ \cite{joint}. In this model, the licensed users which are also the source of interference are not static but rather, highly dynamic. Therefore, we consider that their locations with respect to the reference receiver are constantly changing during each transmission. Considering a Poisson point process $\Phi$, it is possible to characterize the signal-to-interference ratio at the reference link SIR$_0$ as in \cite{ramezanipour2018increasing}. Then, the outage probability $P_\textrm{out}=\textup{Pr}\left[\textup{SIR}_0 \leq \beta\right]$ for each transmission attempt is as presented below where $k=\pi r_0^2 \Gamma{\left(1 - \frac{2}{\alpha} \right)} \Gamma{\left(1 + \frac{2}{\alpha} \right)}$ \cite{ramezanipour2018increasing}.

\begin{equation}\label{eq:6}
P_\textrm{out}= 1 - e^{- k \lambda  \beta^{2/\alpha}},
\end{equation}

We are now able to evaluate the link throughput $T$ in the reference link as:

\begin{equation}\label{eq:5}
T=\frac{\log(1+\beta)}{1+\bar{m}}\left(1-P_\textrm{out}^{1+m}\right),
\end{equation}

\noindent It should be noted that in this throughput equation, $m$ shows the retransmissions attempts whereas $1+\bar{m}$ is the average number of transmissions for a successful transmission. Further details shall be seen in section III. As stated above, $\left(1-P_\textrm{out}^{1+m}\right)$ is the probability of having a successful transmission.


%
\section{Throughput optimization}\label{sect:tput_analysis}
\subsection{Constrained optimization}

In this section we evaluate the optimal link throughput in the sense of spectral efficiency. We consider a constrained throughput optimization problem where the constraint is a maximum acceptable error rate which is imposed by the application at hand. This means that the quality requirement of this network is determined by how often a message is eventually dropped after all retransmission attempts. Considering \eqref{eq:5}, the optimization problem is defined as below.

\begin{equation}
\begin{aligned}\label{eq:7}
& \underset{(\beta,m)}{\text{max}}
& & \frac{\log(1+\beta)}{1+\bar{m}}\times\left(1-P_\textrm{out}^{1+m}\right)  \\ 
& \text{subject to}
& & P_\textrm{out}^{1+m} \leq \epsilon ,
\end{aligned}.
\end{equation} 

\noindent where $\epsilon$ represents the aforementioned quality requirement. In this equation, the SIR threshold $\beta>0$ and the number of allowed retransmissions $m \in \mathbb{N}$ are the design variables.

\begin{lemma}
	The throughput $T$ in \eqref{eq:5} is a function of the variables $m>0$ and $\beta>0$, i.e.  $T = f (\beta, m)$. The function $f$ is then concave with respect to $\beta$ if $\frac{\partial^2 T}{\partial \beta^2} < 0$. After that, we can calculate  $\beta^\ast$ which is the value of the SIR threshold that maximizes the link throughput.
\end{lemma}

\begin{equation}\label{eq:b}
\beta^\ast=\left(- \frac{1}{k\lambda} \log{\left (1 - \epsilon^{\frac{1}{m + 1}} \right )}\right)^{\frac{\alpha}{2}}.
\end{equation}

\begin{IEEEproof}
	As $m$ and $\beta$ are strictly positive variables and function $T$ is twice differentiable in terms of $\beta$, then $T$ is concave if and only if $\frac{\partial^2 T}{\partial \beta^2} < 0$. Based on our calculations, we can see below that the second derivative of the throughput equation is in fact negative with respect to $\beta$. Hence, in the region $\frac{\partial^2 T}{\partial \beta^2} < 0$, $T$ is concave.
	
	\begin{equation}
	\label{eq:d}
	\frac{\partial^2 T}{\partial \beta^2}=- \frac{\left(- \epsilon + 1\right) \left(- \epsilon^{\frac{1}{m + 1}} + 1\right) \left(- \epsilon^{m + 1} + 1\right)}{\left(\beta + 1\right)^{2} \left(- \epsilon \left(- \epsilon^{\frac{1}{m + 1}} + 1\right) \left(m + 1\right) - \epsilon + 1\right)},
	\end{equation}
	
Knowing this, we can now calculate  $\beta^\ast$. To do so, we follow the same steps as in \cite[Prop.1]{nardelli2016maximizing} where $\beta^\ast$ is the highest values that satisfies the inequality $1-P_{suc}=\epsilon$, thus $\beta^\ast$ which is presented as \eqref{eq:b}.


With this result at hand, we move forward as follows. The constraint in the optimization problem is $ P_\textrm{out}^{1+m} \leq \epsilon $. By considering the equality part $P_\textrm{out}^{1+m} = \epsilon$ in addition to  \cite[§17]{nardelli2009multi}, we reach the below equation for the average number of retransmission attempts $1+\bar{m}$:

\begin{equation}\label{eq:4}
1+\bar{m}= \frac{- \epsilon \left(- \epsilon^{\frac{1}{m + 1}} + 1\right) \left(m + 1\right) - \epsilon + 1}{\left(- \epsilon + 1\right) \left(- \epsilon^{\frac{1}{m + 1}} + 1\right)},
\end{equation}

\noindent hence, by inserting \eqref{eq:6} and \eqref{eq:4} into \eqref{eq:5}, we reach the following as the throughput equation.

\begin{equation}\label{eq:t}
T =\log{\left (\beta + 1 \right )} \frac{\left(- \epsilon + 1\right) \left(- \epsilon^{\frac{1}{m + 1}} + 1\right) \left(- \epsilon^{m + 1} + 1\right) }{- \epsilon \left(- \epsilon^{\frac{1}{m + 1}} + 1\right) \left(m + 1\right) - \epsilon + 1}. 
\end{equation}

\end{IEEEproof}
\begin{figure*}[!t]
	\begin{align}\label{eq:mast}
	m^\ast=  \max\limits_{m \in \mathbb{N}}  & \;\; \frac{\alpha \epsilon^{\frac{1}{m + 1}} \left(- \frac{1}{k} \log{\left (- \epsilon^{\frac{1}{m + 1}} + 1 \right )}\right)^{\frac{\alpha}{2}} \left(- \epsilon + 1\right) \left(- \epsilon^{m + 1} + 1\right) \log{\left (\epsilon \right )}}{2 \left(m + 1\right)^{2} \left(\left(- \frac{1}{k} \log{\left (- \epsilon^{\frac{1}{m + 1}} + 1 \right )}\right)^{\frac{\alpha}{2}} + 1\right) \left(- \epsilon \left(- \epsilon^{\frac{1}{m + 1}} + 1\right) \left(m + 1\right) - \epsilon + 1\right) \log{\left (- \epsilon^{\frac{1}{m + 1}} + 1 \right )}}\nonumber\\
	& + \frac{\epsilon^{\frac{1}{m + 1}} \left(- \epsilon + 1\right) \left(- \epsilon^{m + 1} + 1\right) \log{\left (\epsilon \right )} \log{\left (\left(- \frac{1}{k} \log{\left (- \epsilon^{\frac{1}{m + 1}} + 1 \right )}\right)^{\frac{\alpha}{2}} + 1 \right )}}
	{\left(m + 1\right)^{2} \left(- \epsilon \left(- \epsilon^{\frac{1}{m + 1}} + 1\right) \left(m + 1\right) - \epsilon + 1\right)} \nonumber\\
	&- \frac{\epsilon^{m + 1} \log{\left (\epsilon \right )} \log{\left (\left(- \frac{1}{k} \log{\left (- \epsilon^{\frac{1}{m + 1}} + 1 \right )}\right)^{\frac{\alpha}{2}} + 1 \right )}}{- \epsilon \left(- \epsilon^{\frac{1}{m + 1}} + 1\right) \left(m + 1\right) - \epsilon + 1} \left(- \epsilon + 1\right) \left(- \epsilon^{\frac{1}{m + 1}} + 1\right)\nonumber\\
	& + \frac{\log{\left (\left(- \frac{1}{k} \log{\left (- \epsilon^{\frac{1}{m + 1}} + 1 \right )}\right)^{\frac{\alpha}{2}} + 1 \right )}}{\left(- \epsilon \left(- \epsilon^{\frac{1}{m + 1}} + 1\right) \left(m + 1\right) - \epsilon + 1\right)^{2}} \left(- \epsilon + 1\right) \left(- \epsilon^{\frac{1}{m + 1}} + 1\right) \left(- \epsilon^{m + 1} + 1\right) \left(\frac{\epsilon \epsilon^{\frac{1}{m + 1}}}{m + 1} \log{\left (\epsilon \right )} + \epsilon \left(- \epsilon^{\frac{1}{m + 1}} + 1\right)\right)
	\end{align}
	\hrule 
\end{figure*}

\begin{proposition}
	The maximum allowed number of retransmissions $m^\ast$ that maximizes the link throughput is then given by \eqref{eq:mast}.
\end{proposition}

\begin{IEEEproof}
	By taking into account the $\beta^\ast$ from (\ref{eq:b}), and (\ref{eq:t}), we can find the optimal maximum number of retransmissions ($m^\ast$) for the throughput.
	%
	Hence, the optimal throughput $T^*$ in terms of both $m$ and $\beta$ is then given by the value of $m$ that maximizes the throughput, which achieved by \eqref{eq:mast}.  It should be noted that the maximum number of retransmissions $m$ is a natural number that is usually small, therefore, \eqref{eq:mast} is easy to evaluate.  	
\end{IEEEproof}

\subsection{Extreme Cases}

In this section, we evaluate the two extreme cases in terms of maximum number of transmissions and their effect on the throughput. To do so, we consider having no transmission $m=0$ and having a very high number of retransmissions $m \rightarrow \infty$ in the network. Considering the zero transmission case, we reach the following for the throughput:

	\begin{align}\label{eq:x}
	T &= \log(1+\beta) \left(1 - \epsilon^{\frac{1}{m + 1}}\right) \nonumber\\
	&= \log(1+\beta) \left(1 - P_\textrm{out}\right)\nonumber\\
	&= \log(1+\beta) e^{- k \lambda  \beta^{2/\alpha}}.
	\end{align}

Moving on to the $m \rightarrow \infty$ case, by replacing $\epsilon=P_\textrm{out}^{(1+m)}$ in \eqref{eq:4}, we reach the below equation for the average number of retransmission for the extreme cases ($1+\bar{m}_e$):

\begin{equation}\label{m2}
1+\bar{m}_e= \frac{P_\textrm{out}^{(1+m)}(1+m)(P_\textrm{out}-1)-P_\textrm{out}^{(1+m)}+1}{(P_\textrm{out}-1)(P_\textrm{out}^{(1+m)}-1)},
\end{equation}

\noindent now by inserting \eqref{m2} in \eqref{eq:5}, we can achieve the throughput equation in this case. After some mathematical manipulation, it is proven that while $m \rightarrow \infty$, we reach the same throughput equation as in \eqref{eq:x}. Hence, for both extreme cases of $m=0$ and $m \rightarrow \infty$, the system behavior remains the same in terms of throughput. We later use these results in the numerical results section of this paper and discuss what they show and why they are important. More details about these two extreme cases can also be found in \cite{haenggi2013local}.

\begin{figure}[t]
	\centering
	\includegraphics[width=\columnwidth]{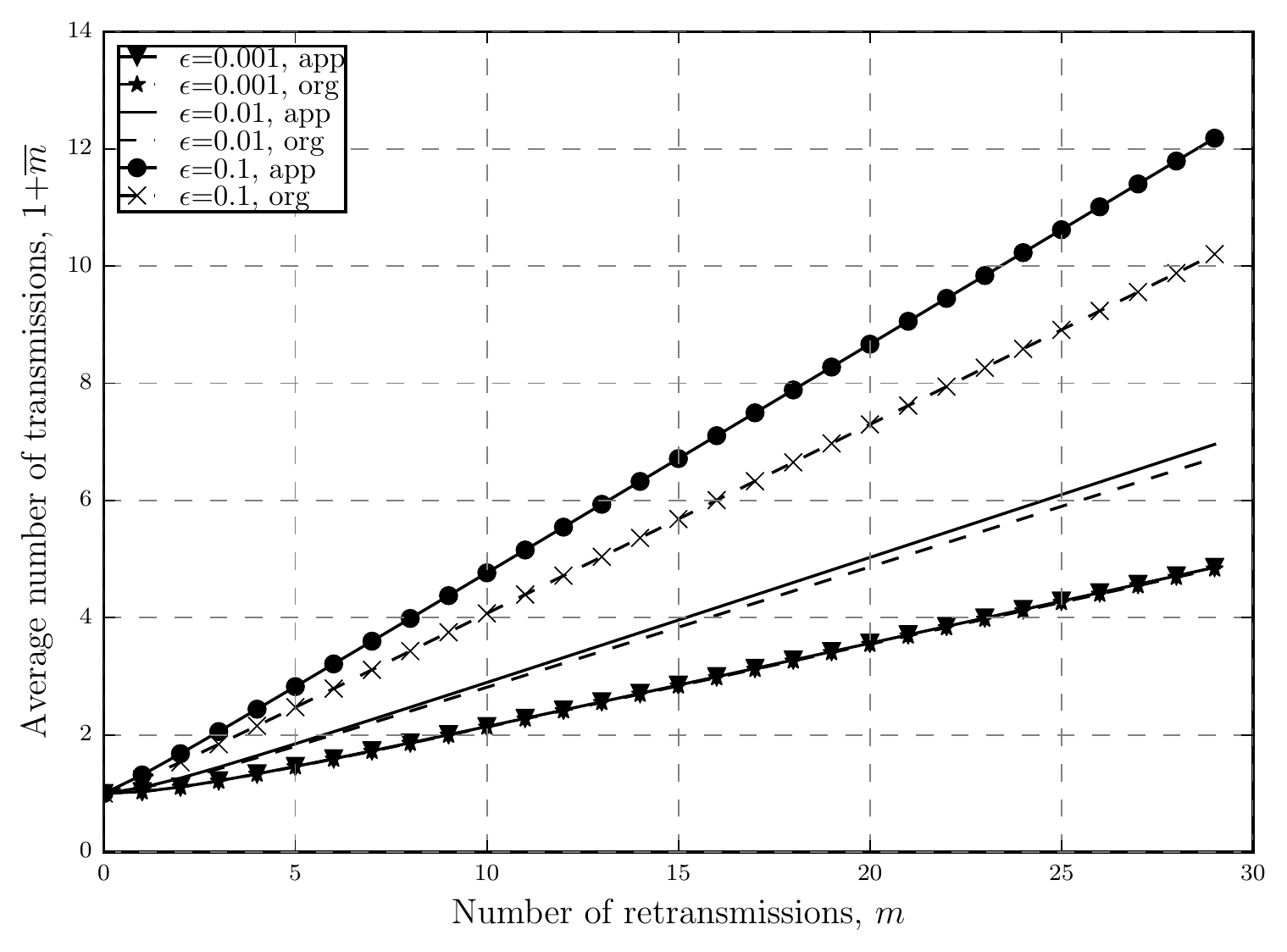}
	\caption{Comparison between the approximated and original average number of retransmissions $1+\bar{m}$ versus the number of retransmissions with different outage threshold $\epsilon$ levels.}
	\label{fig:E1}
\end{figure}
\subsection{Error Analysis}

\begin{figure}[t]
	\centering
	\includegraphics[width=\columnwidth]{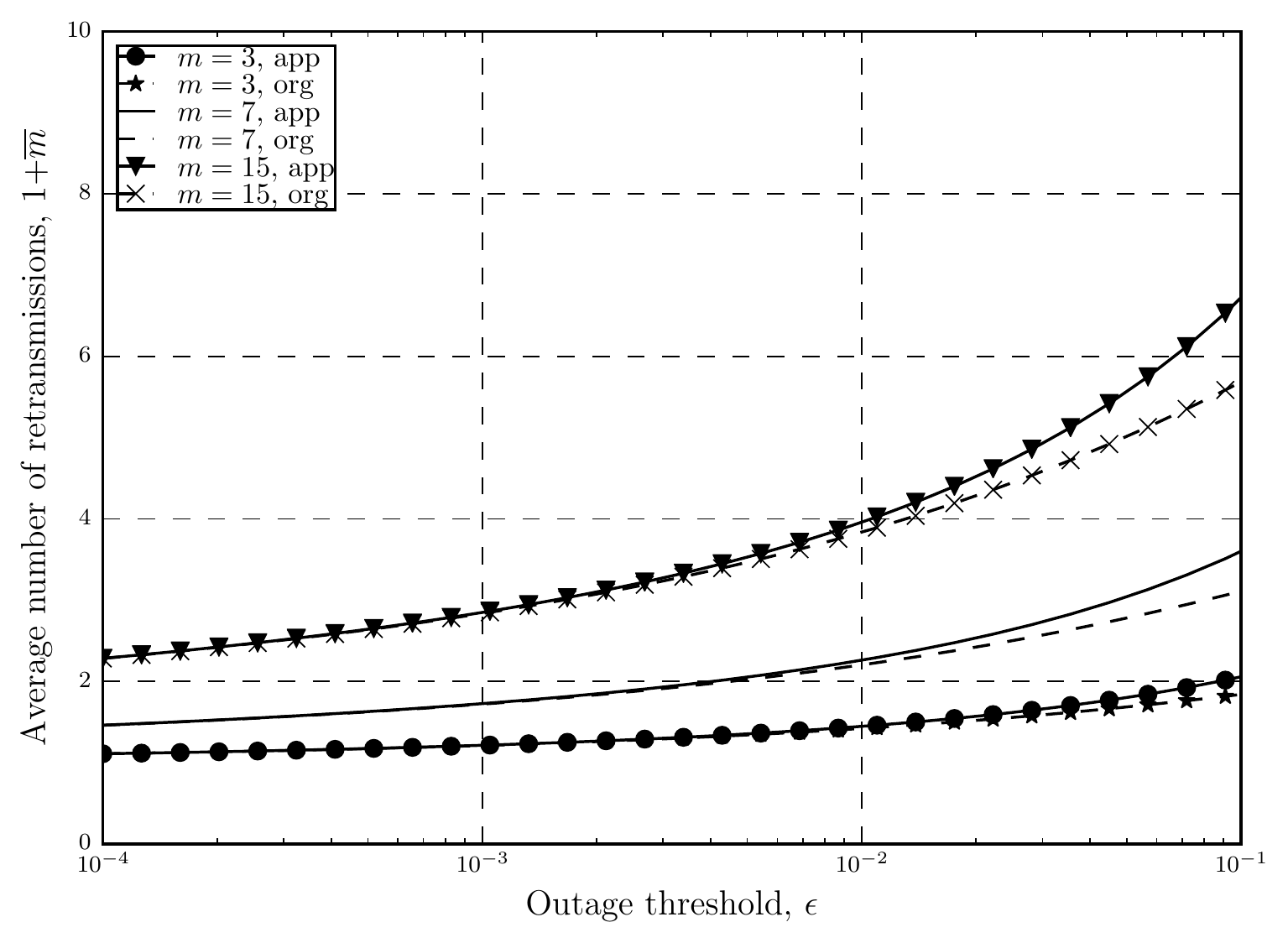}
	\caption{Comparison between the approximated and original average number of retransmissions $1+\bar{m}$ versus the outage threshold $\epsilon$ with different number of retransmissions.}
	\label{fig:E2}
\end{figure}

In our previous work, we can see that in \cite[Lemma 2]{ramezanipour2018increasing}, we use an approximation of \eqref{eq:4} (which is also shown in this paper as \eqref{eq:old}) presented in this analysis in order to calculate the average number of retransmissions in the throughput optimization problem in \cite{ramezanipour2018increasing}. In this section we aim to prove that this previously used approximation is in fact a tight and good approximation and the error in calculations is low when using the approximated expression. It is important to remember that this approximation is suitable when the number of transmissions is not not large and also $\epsilon <  40\%$ which is the case if most of practical communications anyway since having a very large $m$ is not efficient and $ 40\%$ is already a quite a loose error threshold for most systems.

In Figs. \ref{fig:E1} and \ref{fig:E2}, we can see how \eqref{eq:4} and \cite[§6]{ramezanipour2018increasing} compare to each other with respect to increasing the number of retransmissions and outage threshold respectively. From these figures, we confirm our previous claims and show the accuracy of the approximated expression. In these figures, the previously mentioned fact about \cite[§6]{ramezanipour2018increasing} being suitable for low $m$ and $\epsilon$ is also proven. It is shown that in both figures, the approximated and the original $1+\bar{m}$ are either the same or very close to each other in the mentioned areas.

\begin{figure}[!t]
	\centering
	\includegraphics[width=\columnwidth]{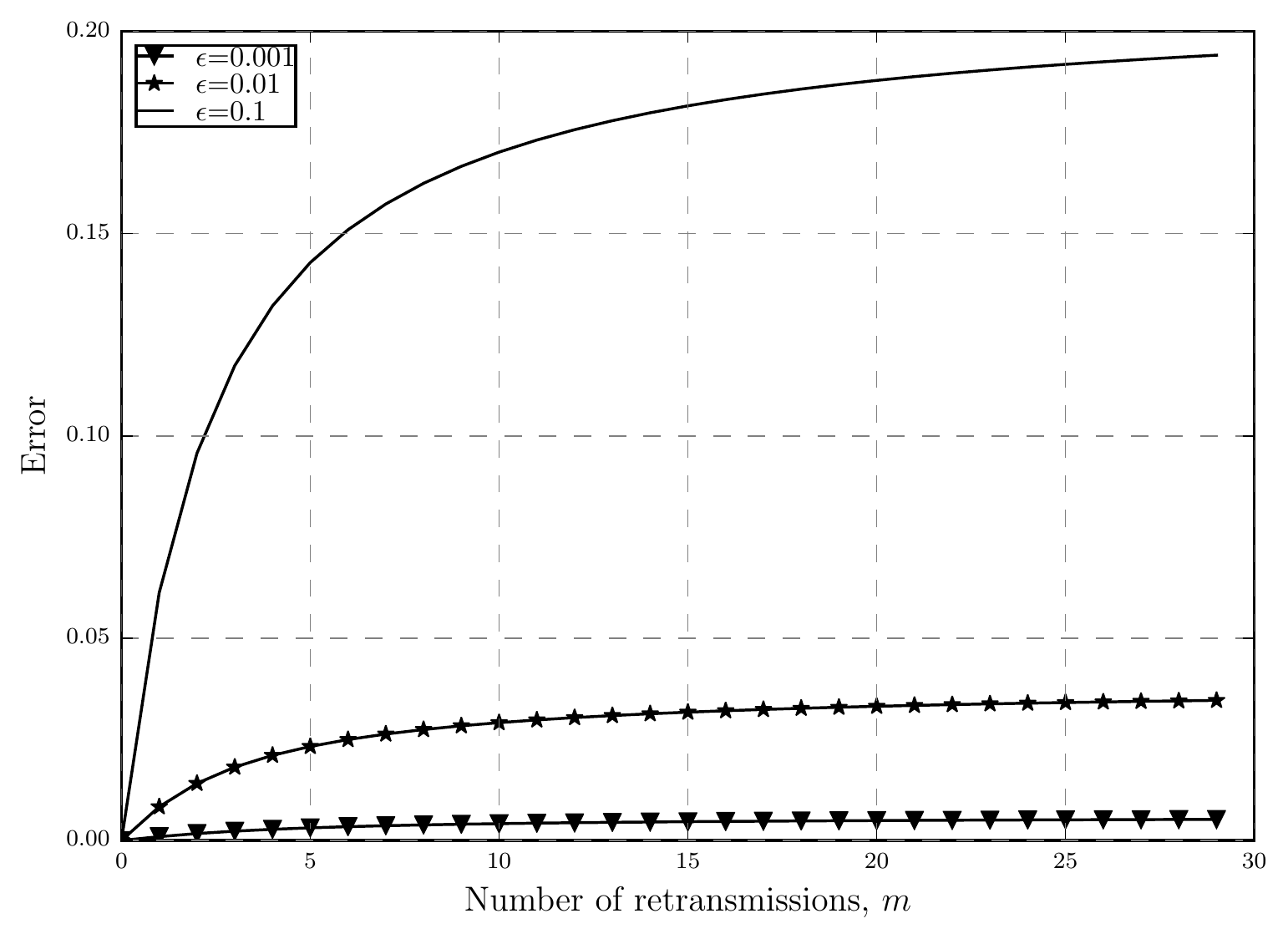}
	\caption{Error between the approximated and original average number of retransmissions $1+\bar{m}$ versus  the number of retransmissions with different outage threshold $\epsilon$ levels.}
	\label{fig:E3}
\end{figure}

\begin{figure}[!t]
	\centering
	\includegraphics[width=\columnwidth]{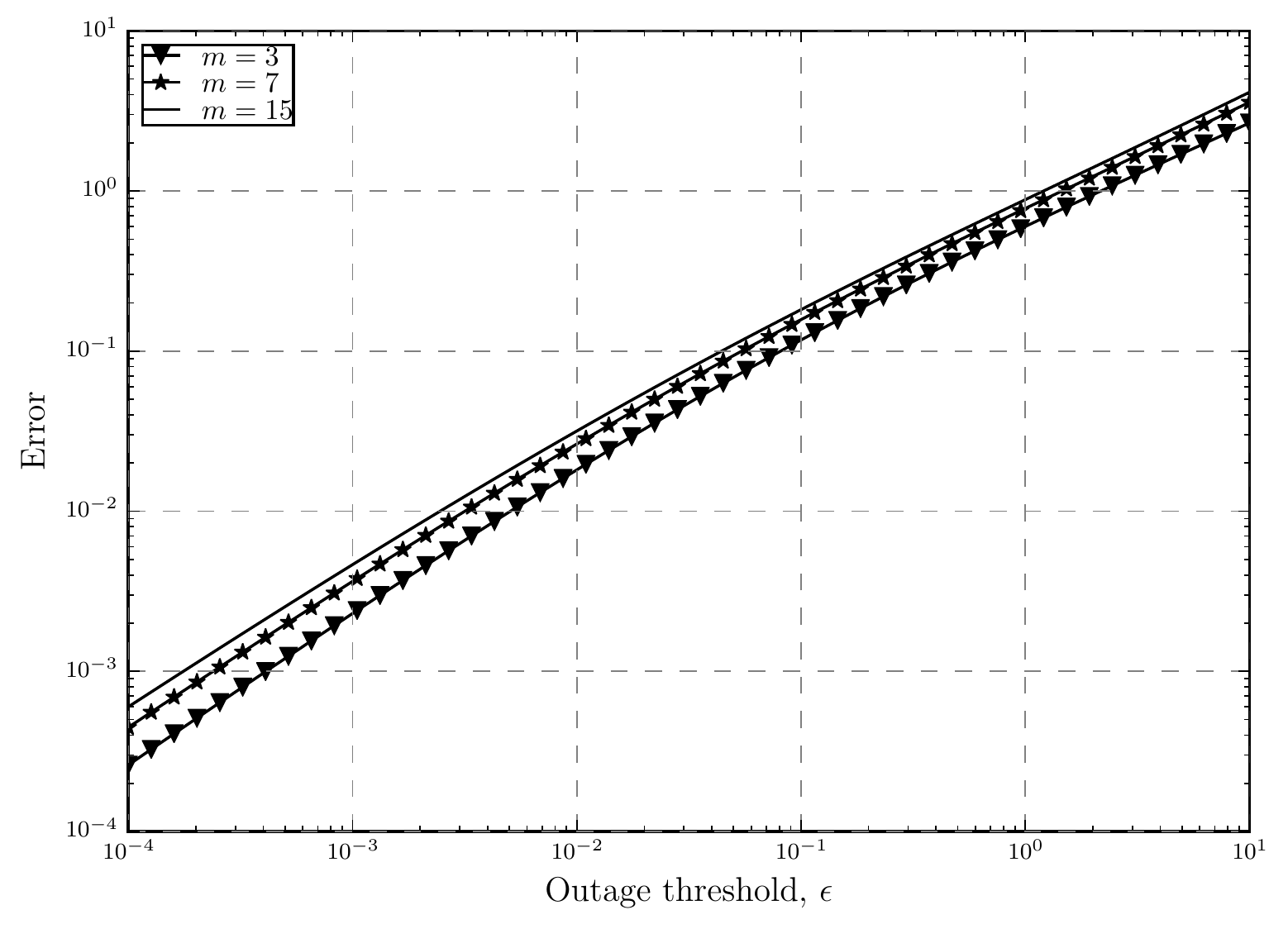}
	\caption{Error between the approximated and original average number of retransmissions $1+\bar{m}$ versus  the number of retransmissions with different outage threshold $\epsilon$ levels.}
	\label{fig:E4}
\end{figure}
 Moving on, we analyze the error between $1+\bar{m}_{app}$ and $1+\bar{m}_{org}$ as:
 
 \begin{equation}\label{eq:old}
 1+\bar{m}_{app}= \sum_{n=0}^{m} \; P_\textrm{out}^n \approx \frac{1-P_\textrm{out}^{1+m}}{1-P_\textrm{out}} \approx \frac{1-\epsilon}{1-\epsilon^{\frac{1}{1+m}}}.
 \end{equation}
 
\begin{equation}\label{e}
Error=\abs {\frac{(1+\bar{m}_{app})-(1+\bar{m}_{org})}{(1+\bar{m}_{org})}}\,
\end{equation}
 
\noindent where $(1+\bar{m}_{app})$ is  \cite[§6]{ramezanipour2018increasing} and $(1+\bar{m}_{org})$ is \eqref{eq:4}. Fig. \ref{fig:E3} shows this error as a function of number of retransmissions while Fig. \ref{fig:E4} shows how the error changes as a function of the outage threshold. We can see that as expected, the error increases when $m$ or $\epsilon$ get larger. It is also shown that the increase in $\epsilon$ has a bigger impact on increasing the error. In Fig. \ref{fig:E3}, when we have strict outage threshold in the network, increasing $m$ does not increase the error much. However, as the outage threshold gets looser, increasing $m$ results in higher error and we can see that for $\epsilon=0.1$, error can even reach $20\%$. This is also true for Fig. \ref{fig:E4} where the highest error happens when both $m$ and $\epsilon$ are high, further proving the point that the approximation is tight only for low  $m$ and $\epsilon$.

\section{Energy Efficiency Optimization}

After the throughput optimization, we are now moving on to the energy efficiency (EE) optimization problem. This is important specially since the energy efficiency can be seen as a tool that represents the trade-off between the throughput and total power consumption ($\textup{P}_{T}$) in a network. The total power consumption in the network is itself a function of the distance dependent transmission power, total energy consumed by the radio components and bit rate \cite{cui2005energy, i1,de2011energy, alves2014outage}. Having this in mind, the total power consumption of this model is calculated as:

\begin{equation}\label{ee}
\textup{P}_{T}=\sum_{1}^{m+1} \frac{P_{PA}+P_{T_x}+P_{R_x}}{\textup{log} (1+\beta^*)}.
\end{equation}

In the above equation, $P_{PA}$ is the energy consumed by the power amplifier in an one-hop communication network. This consumed energy is itself a function of the drain efficiency parameter of the amplifier. This parameter is shown by $\delta$ and is $\delta=0.35$, hence, $P_{PA}=\frac{\beta^*}{\delta}$. As it was also mentioned earlier, $\beta^\ast$ is the optimal SIR threshold and $\textup{log} (1+\beta^*)$ represents the bit rate (bits/s) of the network. $P_{T_x}$ is the power consumed for the transmission, which is constant and $P_{T_x}=97.9$ mW while $P_{R_x}$ is the consumed energy during reception and is $P_{T_x}=112.2$ mW \cite{de2011energy}. It should be noted that both of these parameters are constant since their value depends on the current technology and depend on the internal circuitry power consumption. Thus, we can now express the energy efficiency as:

\begin{equation}\label{eq:10}
\textup{EE}=\frac{T}{\textup{P}_{T}}=\frac{\textup{log}(1+\beta)(1-\epsilon^{(1+m)})}{(1+\bar{m})(\frac{\beta^*}{\delta}+\textup{P}_{T})},
\end{equation}

The energy efficiency optimization problem is then :

\begin{equation}
\begin{aligned}\label{eq:ee1}
& \underset{(\beta,m)}{\text{max}}
& & \frac{\textup{log}(1+\beta)(1-\epsilon^{(1+m)})}{(1+\bar{m})(\frac{\beta^*}{\delta}+\textup{P}_{T})} \\ 
& \text{subject to}
& & P_\textrm{out}^{1+m} \leq \epsilon 
\end{aligned},
\end{equation}

\noindent which like throughput, is also a function of SIR threshold $\beta>0$ and the number of allowed retransmissions $m \in \mathbb{N}$. The energy efficiency equation in \eqref{eq:ee1} is concave with respect to $\beta$ if $\frac{\partial^2 EE}{\partial \beta^2} < 0$ and \eqref{eq:ee2} obtained for $\frac{\partial^2 EE}{\partial \beta^2}$ proves that this in fact is true and the energy efficiency is concave with respect to $\beta$. Energy efficiency is also concave with respect to $m$ but since the obtained expression is long and complicated, we show this concavity and the optimal throughput in the numerical results section of this paper.

\begin{figure*}[!t]
	\begin{align}\label{eq:ee2}
	\frac{\partial^2 EE}{\partial \beta^2}=- \frac{\left(- \epsilon + 1\right) \left(- \epsilon^{\frac{1}{m + 1}} + 1\right) \left(- \epsilon^{m + 1} + 1\right)}{\left(\beta + 1\right)^{2} \left(pc + \frac{1}{\eta} \left(- \frac{1}{k} \log{\left (- \epsilon^{\frac{1}{m + 1}} + 1 \right )}\right)^{\frac{\alpha}{2}}\right) \left(- \epsilon \left(- \epsilon^{\frac{1}{m + 1}} + 1\right) \left(m + 1\right) - \epsilon + 1\right)}
	\end{align}
	\hrule 
\end{figure*}

\section{Numerical Results}

In this section, we present the numerical results for the previously studied optimal throughput $T^*$ and optimal energy efficiency $EE^*$. It is important to mention that to obtain these results, the following arbitrary parameters where considered in our simulations. The distance between the reference receiver and sensor $r_0=1$ and path-loss exponent $\alpha=4$; the required error rate $\epsilon$ and the density of interferers $\lambda$ are the input parameters that their effects are analyzed. Moreover, based on \cite{de2011energy}, $P_{T_x}=97.9$ mW, $P_{R_x}=112.2$ mW and $\delta=0.35$ were considered.

\begin{figure}[t]
	\centering
	\includegraphics[width=\columnwidth]{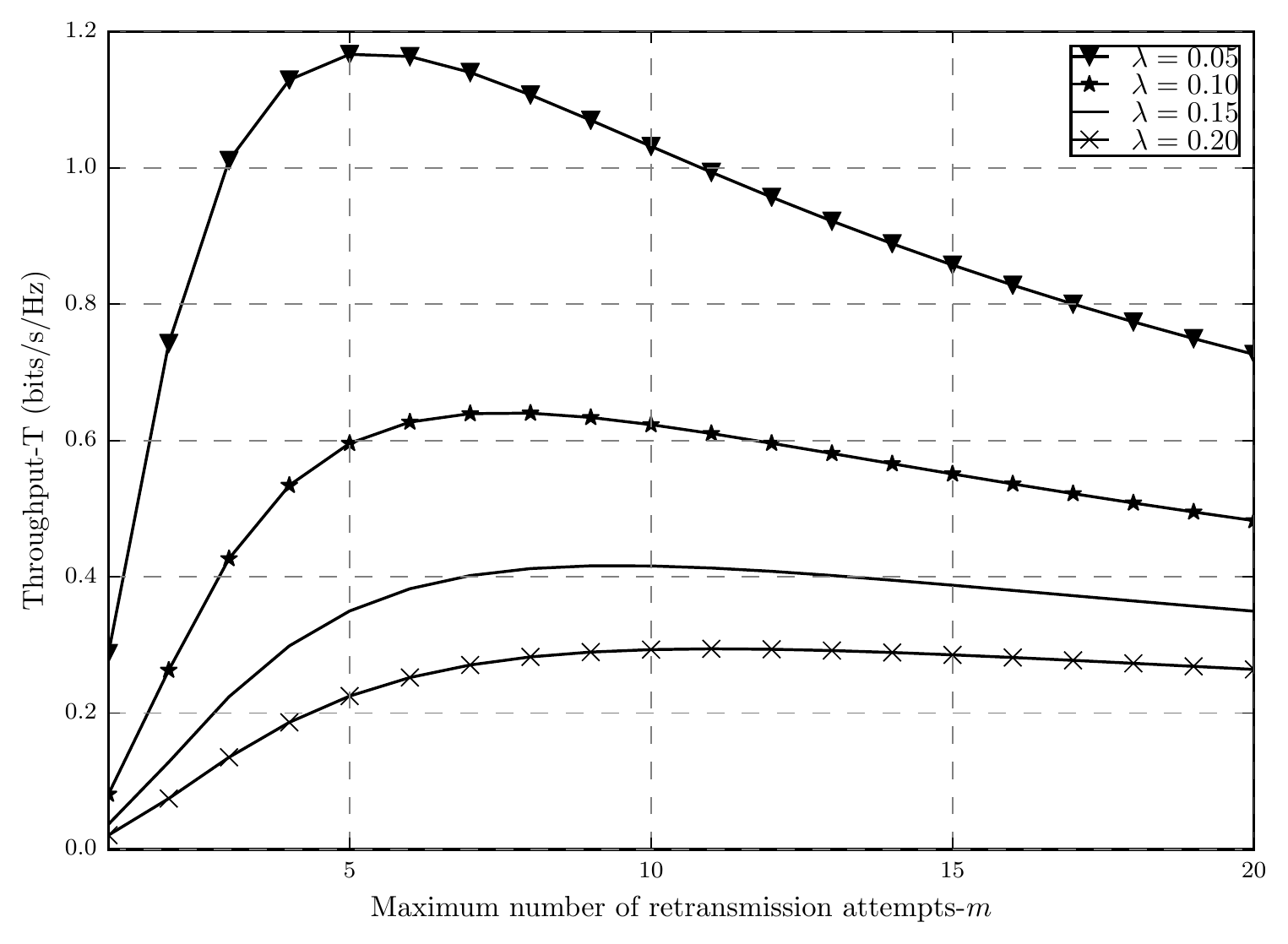}
	\caption{Throughput $T$ versus the maximum number of allowed retransmissions attempts $m$ for $\alpha=4$, $r_0=1$, $\epsilon=0.02$ and different densities $\lambda$.}
	\label{fig:T1}
\end{figure}

Fig. \ref{fig:T1} shows how the throughput behaves as a function of the maximum number of retransmissions in with different network densities. In this plot, we consider $\epsilon=0.02$ in order to show how tight the previously mentioned approximation is. If Fig. \ref{fig:T1} in this paper is compared with \cite[Fig. 3]{ramezanipour2018increasing}, we can see that the two plots are almost identical since here we have strict outage threshold in the network which is the area that the approximations works well.

\begin{figure}[t]
	\centering
	\includegraphics[width=\columnwidth]{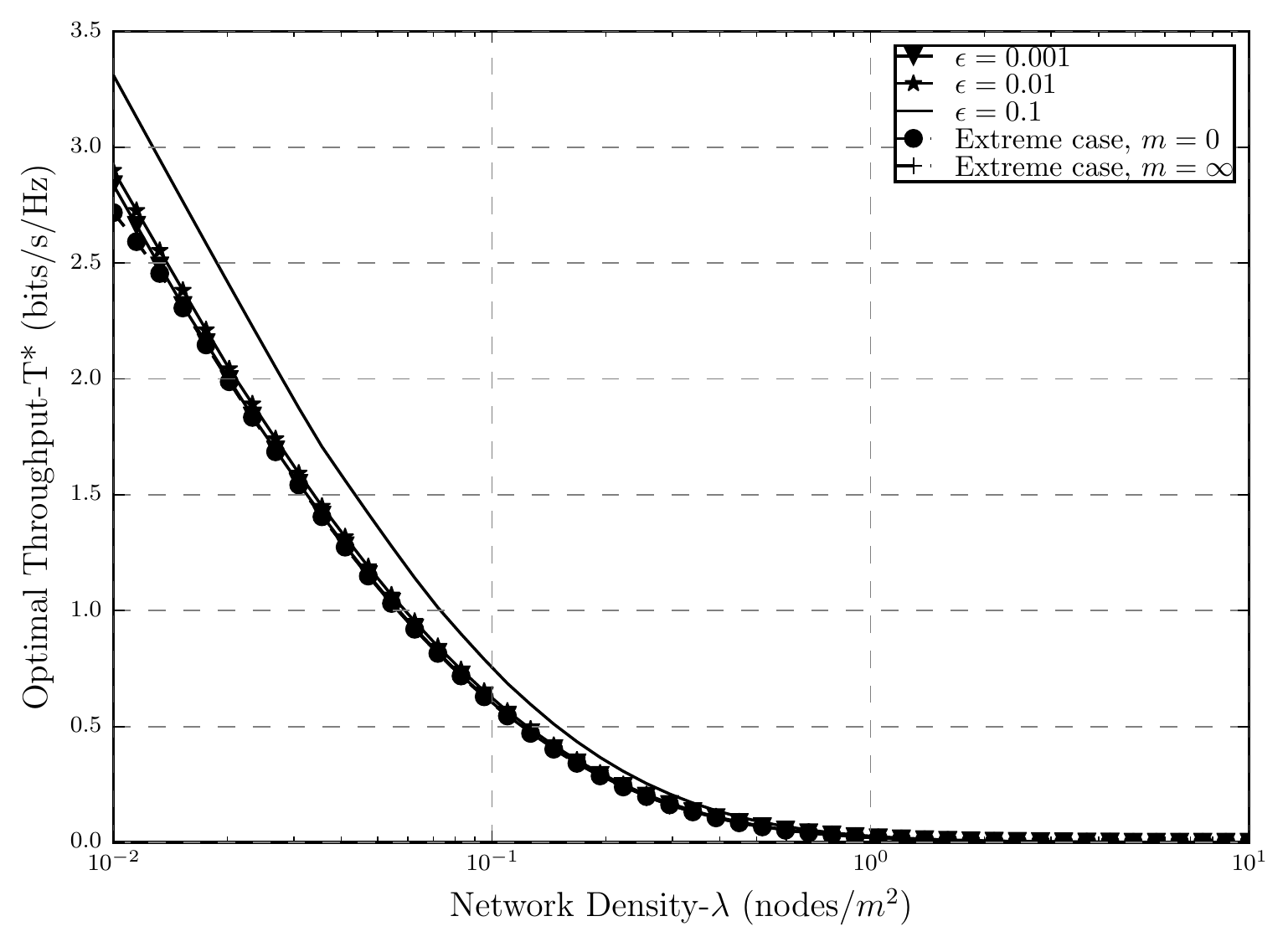}
    \caption{Optimal Throughput $T^\ast$ versus the density of interferers $\lambda$  where $\alpha=4$, $r_0=1$.}
	\label{fig:T2}
\end{figure}

It is shown that in Fig. \ref{fig:T1}, as the number of allowed retransmissions increases, the link throughput also improves until one point at which the throughput stars decreasing. This is true for different network densities as well. By increasing $m$, the system experiences a two fold effect which results in the trade-off. This effect can be explained as with increasing $m$ we are allowing for also higher values of $\beta$ which also means having higher spectral efficiency in each transmission attempt (higher $\textup{log} (1+\beta)$). However, by doing so, we are also increasing the outage probability since this reduces the chance of a message being correctly decoded in a single transmission attempt. In order to capture these trade-offs, we have the studied constrained optimization which its solutions are $m^*$ and $\beta^*$ which results in the optimal throughput $T^*$ which represents the maximum points shown in Fig. \ref{fig:T1}.

Fig. \ref{fig:T2} represents the optimal throughput as a function of the network density for $5$ different cases. First, we can see that for the stricter outage threshold range, the optimal throughput obtained by the optimization problem in this paper, is the same as the one in  \cite[Fig. 4]{ramezanipour2018increasing}, which again proves how precise the approximation is \footnote{ It is important to mention that different scales are used in the two paper for the plots. In \cite[Fig. 4]{ramezanipour2018increasing}, the linear scale is used in order to plot the optimal throughput whereas in Fig.\ref{fig:T2} in this paper, logarithmic is used, hence, the difference seen between the two plots. }. Moreover, we can see how the optimal throughput decreases as $\lambda$ increases. Since $\lambda$ is an indicator of the number of active transmitters, i.e. source of interference, in the network, it is understandable that as it increases, the throughput decreases since the unlicensed network experiences higher level of interference from the licensed nodes.

As it was proven in \ref{sect:tput_analysis}-B, both extreme cases of having zero and very large number of retransmissions will result in the same throughput. This can also be seen in Fig. \ref{fig:T2} that the optimal throughput obtained from both of these cases is also the same. One interesting notion to consider in this plot is that, when the packet error threshold is loose ($\epsilon=0.1$), even though the number of retransmissions is not as large as the $m \rightarrow \infty$, the system outperforms the other cases. In cases with limited transmissions and strict outage threshold ($\epsilon=0.001$, $\epsilon=0.01$), this is understandable because the stricter the error is, the worse the throughput gets and in these cases, we are fixing the number of retransmissions to lead to the outage probability that maximizes the throughput via $\beta$. Compared to the other two extreme cases also, when $m=0$, we are forcing the system to have zero retransmissions in which the $\beta$ would be high and as it was shown, having higher SIR threshold would result in high outage probability as well which will decrease the throughput. As for the $m \rightarrow \infty$, we are optimizing in terms of an error probability of $0$, in oder words, we are basically forcing the packet loss probability to be zero, hence, we are loosing spectral efficiency since infinite retransmissions will use a lot more spectrum resources which is resulting in the worse throughput compared to $\epsilon=0.1$. 

In both of these extreme cases, we are taking the system degree of freedom away in terms of retransmissions and packet loss, which in the first would result in high packet loss probability and in the second in very high delay. On the other hand, in the proposed throughput optimization problem \eqref{eq:7}, where the outage probability and number of retransmissions are the designed variables, we are in fact relaxing the two very tight constraints which were considered in the two extreme cases, hence, giving back the system's degree of freedom of having a certain arbitrary level of outage while also benefiting fro retransmissions if case an outage event happens. The delay of this network would be much lower than the $m\rightarrow 0$ and although the packet loss probability would not be zero, having limited retransmission not also would reduce that, but would also result is the same or even higher system throughput compared to the case where $m=0$ which proves the benefits and importance of the optimized throughput problem studied in this paper.

\begin{figure}[!t]
	\centering
	\includegraphics[width=\columnwidth]{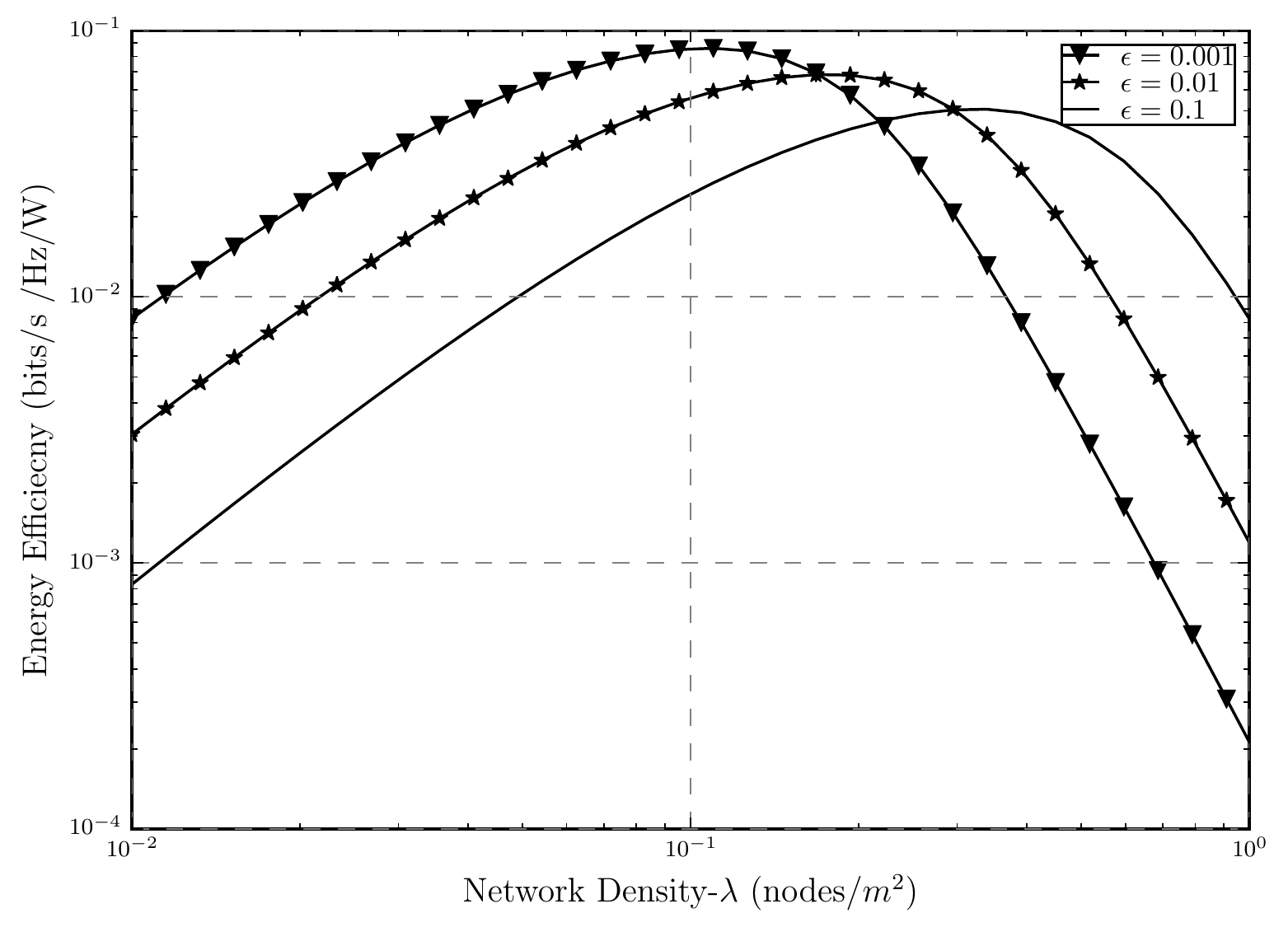}
	\caption{Energy efficiency $EE$ versus the network density $\lambda$ with different outage threshold $\epsilon$ levels and $\alpha=4$, $r_0=1$}
	\label{fig:EE1}
\end{figure}

\begin{figure}[!t]
	\centering
	\includegraphics[width=\columnwidth]{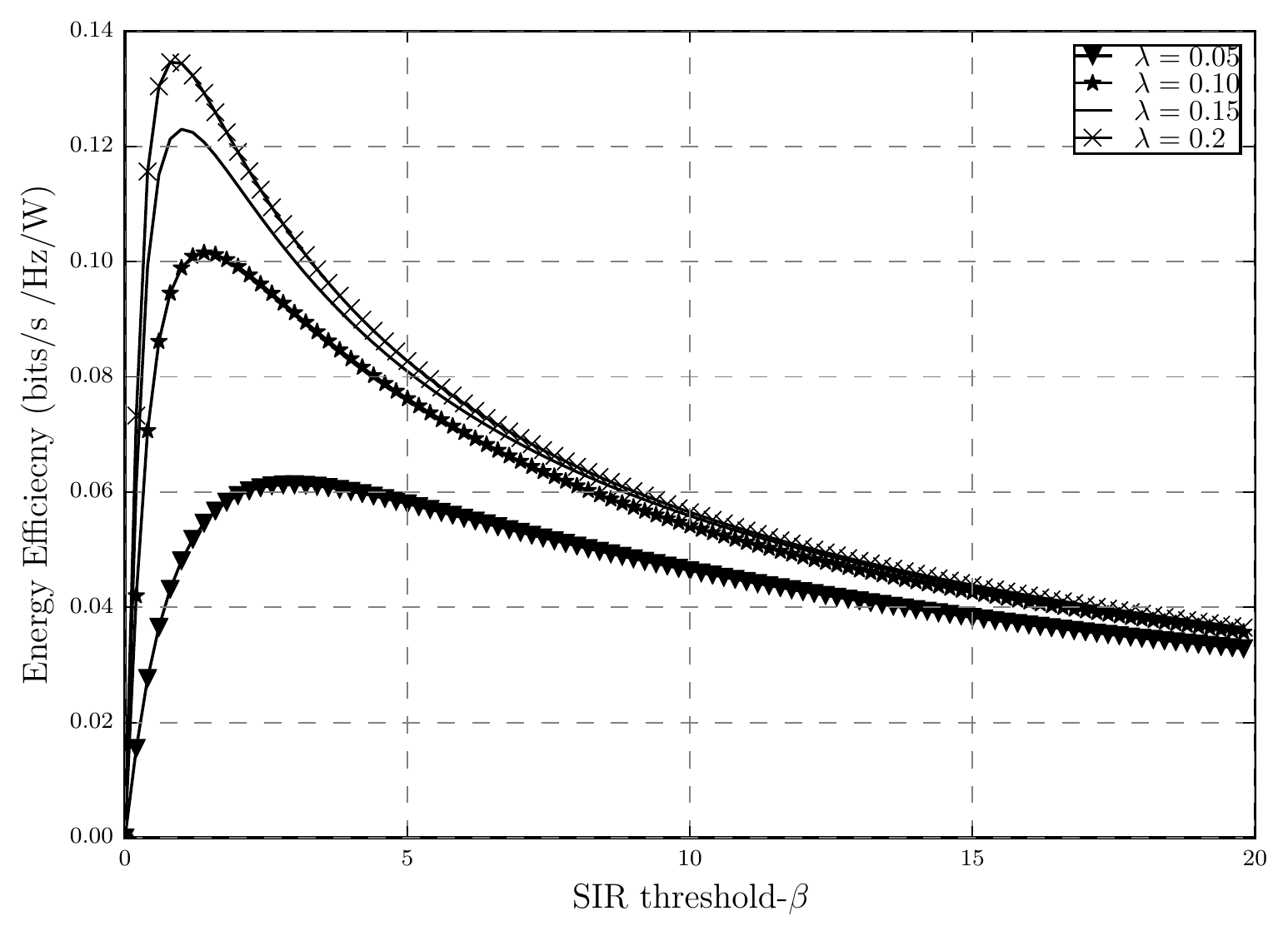}
	\caption{Energy efficiency $EE$ versus the SIR threshold $\beta$ with different network densities $\lambda$ and $\alpha=4$, $r_0=1$, $m=5$ and $\epsilon=0.001$}
	\label{fig:EE2}
\end{figure}

\begin{figure}[t]
	\centering
	\includegraphics[width=\columnwidth]{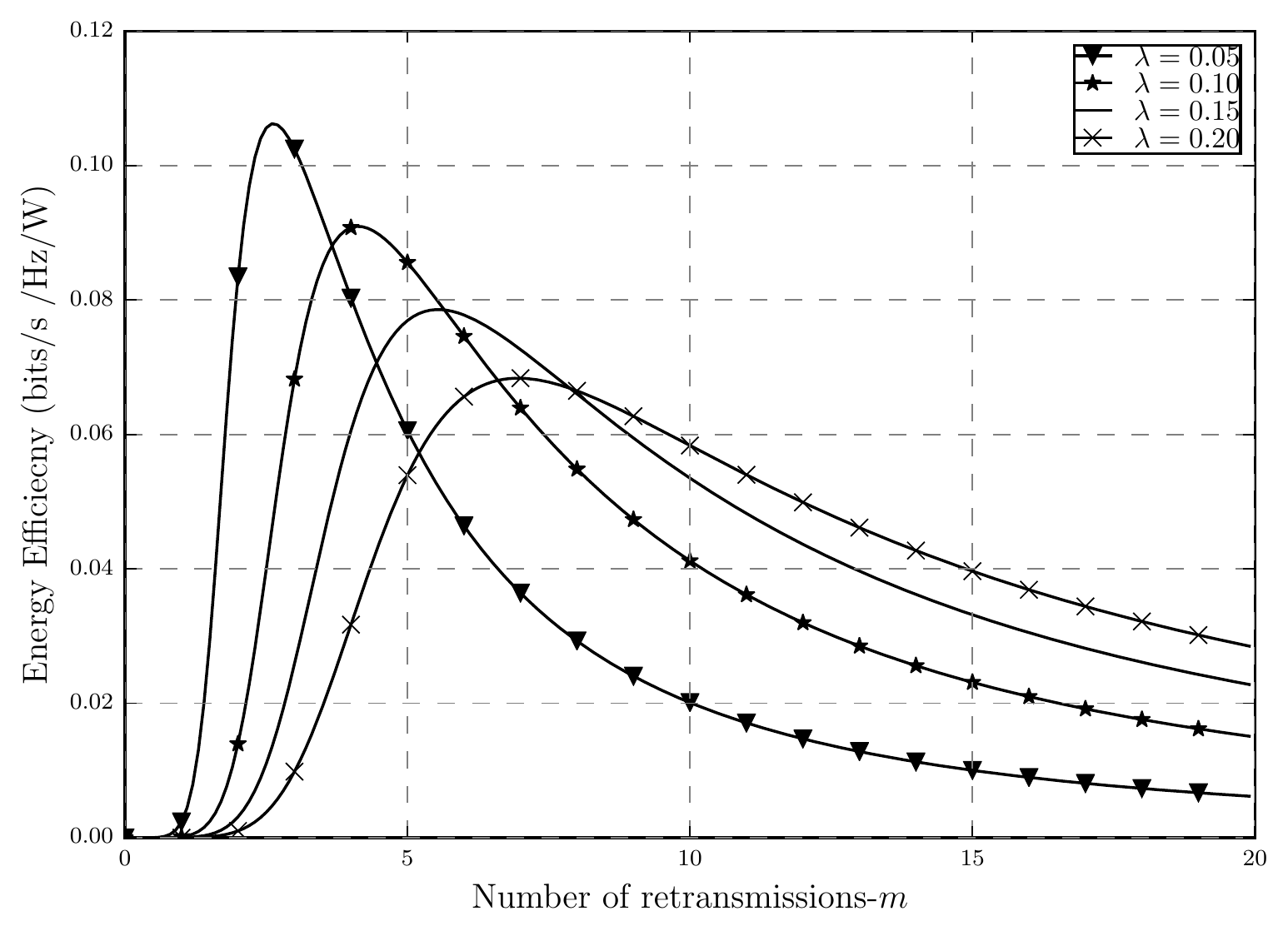}
	\caption{Energy efficiency $EE$ versus the number of retransmissions $m$ with different network densities $\lambda$ and $\alpha=4$, $r_0=1$ and $\epsilon=0.001$}
	\label{fig:EE3}
\end{figure}

Moving on to the energy efficiency analysis, EE is shown in Figs. \ref{fig:EE1}, \ref{fig:EE2} and \ref{fig:EE3} as a function of network density, SIR threshold and number of retransmissions respectively. As can be seen in Fig. \ref{fig:EE1}, EE faces a decrease after some point when $\lambda$ increases. When the nodes are sparsely located in the network, the level of interference is low  but a lot of energy is also used to transfer a message between nodes, hence EE is low. As $\lambda$ gets higher, meaning that the nodes are getting closer to each other, the EE improves since less energy is used for the transmission while interference is still low and the transmission is affected only by path loss. However, when the network gets very dense, the level of interference gets so high that in order to prevent outage, a lot of energy should again be used for transmissions which will cause the decrease in energy efficiency after some point in the plot.

It should be noted that when $\lambda$ is low, the scenario with the loosest outage threshold has the lowest energy efficiency because the SIR in this case is lower compared to the other cases, however, when the network gets denser, the interference level rises which would eventually decrease the throughput, hence, the loosest outage threshold would result in the highest energy efficiency since there is more room allowed for having outage and less energy is used to do the transmission in the presence of the interference. Although having a denser network also means having higher level of interference, when there is room for higher levels of outage, it means that less retransmissions are also needed in order to meet the reliability requirements of the network, hence, less transmission energy is used. All of these would eventually result in the system being more energy efficient in the presence of loose outage threshold in a dense network.

Fig. \ref{fig:EE2} shows how EE performs as the SIR threshold gets larger while having different network densities. As $\beta$ increases, it means that the throughput is also increasing, even though power consumption is also increasing at the same time, the rate at which the throughput is increasing is higher, thus, the energy efficiency also improves during this time. This however means that the outage events are also increasing which will decrease the throughput and eventually EE also decreases. That is why we see the maximum point in Fig. \ref{fig:EE2}. Also, we can see that the denser the network is, the more energy efficient it is. This is due to the fact that while nodes being close to each other means higher interference, it also means less energy is used for the transmission since the distances are smaller in denser networks. This figure also shows that the energy efficiency is concave in terms of $\beta$ which was also proven in the EE optimization problem.

Moreover, the effect of increasing the number of retransmissions on EE with different network densities can be seen in Fig. \ref{fig:EE3}. While $\lambda$ is low, lower numbers of retransmission are needed for a successful message delivery, therefore, the energy efficiency is higher in less dense networks. As the number of retransmissions increases, we can see that while the over all EE starts too decrease, denser networks become more energy efficient as well. The reason behind this is that in networks with high $\lambda$, less energy is used per each retransmission attempt to send the message since the distance between the nodes are smaller. So although the interference is higher, using less transmit power makes the network more energy efficient. 

\begin{figure}[t]
	\centering
	\includegraphics[width=\columnwidth]{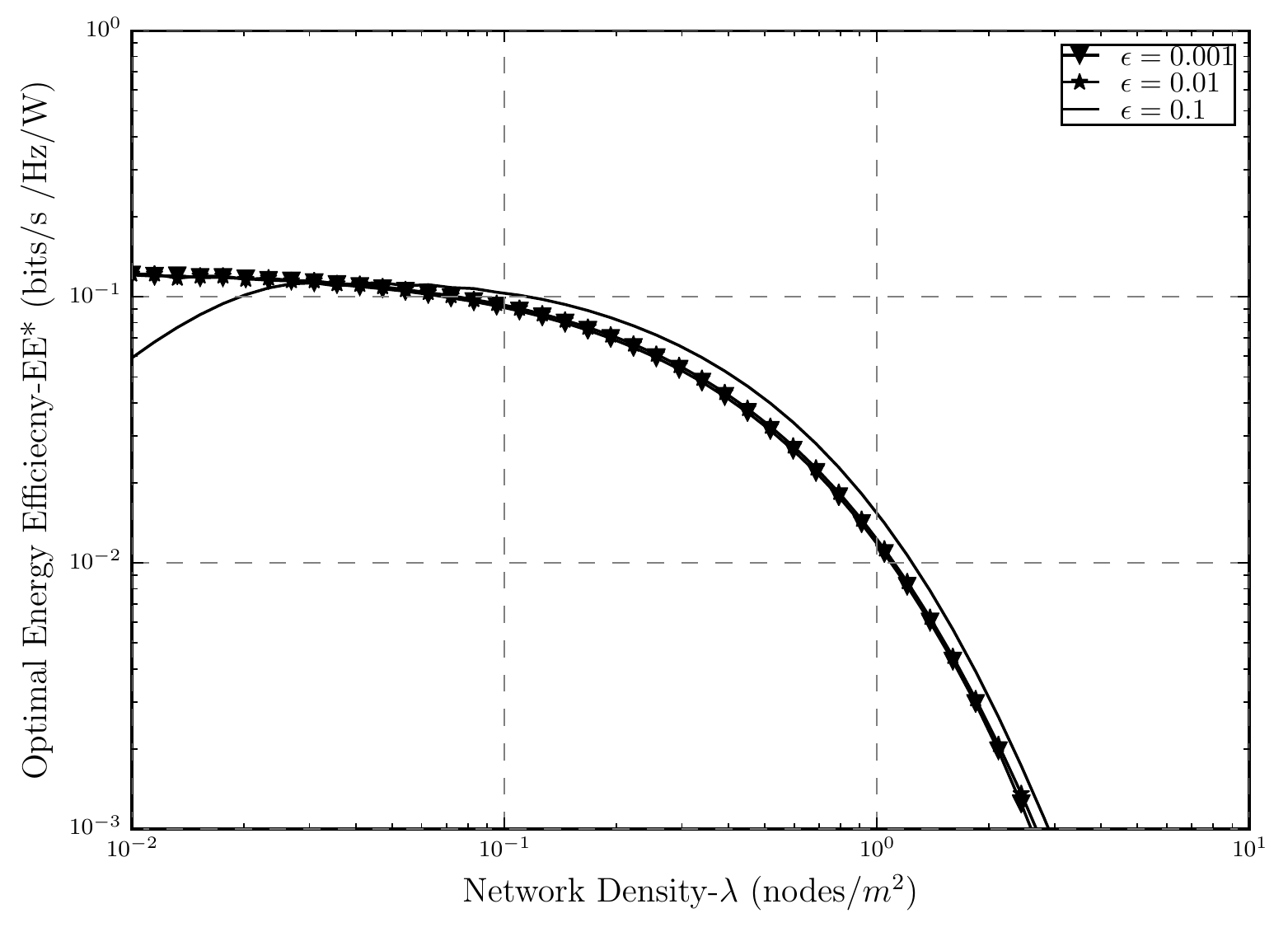}
	\caption{Optimal energy efficiency $EE^{\ast}$ versus the number of retransmissions $m$ with different network densities $\lambda$, $\alpha=4$ and $r_0=1$.}
	\label{fig:EE4}
\end{figure}

In Fig. \ref{fig:EE4} we can see the behavior of the optimal energy efficiency while $\lambda$ and outage threshold level increases. As it was shown earlier, due to the previously explained reasons, the energy efficiency is concave with respect to $\lambda$ meaning that while increasing at first, it starts decreasing after a certain density of interferers is met in the network. Hence, it is understandable that the same thing is happening in the case of optimal energy efficiency as well. It is also shown in this figure that while the scenarios when the outage threshold is somewhat strict in the network ($\epsilon=0.001$, $\epsilon=0.001$), $EE^*$ is almost the same and the difference is negligible. However, when the outage threshold gets loose ($\epsilon=0.1$) the optimal throughput that the system can attain is slightly higher than the other two cases. The crossing point between these cases can also be explained the same way as described in Fig. \ref{fig:EE1} 

\section{Discussion and final remarks}\label{sect:conclusion}

In this paper, we studied the throughput and energy efficiency of a network where licensed and unlicensed nodes share the same frequency band which is used by the licensed nodes for their transmission. The interference in this model comes from the licensed network on the unlicensed nodes. There is also possibility of retransmissions in case the transmitted message in not successfully decoded at the receiver. We studied the optimal throughput in a constrained optimization problems where the interferers locations are modeled using Poisson point process. We then derived the value of optimal number of retransmissions and SIR threshold that jointly result in the optimal throughput. We showed how increasing the number of nodes and the outage threshold can decrease and increase the optimal throughput respectively and how these cases compared to the two extreme cases of having zero and infinite number of retransmissions in the network. We also studied the optimal energy efficiency and how it also decreases with increasing the number of nodes and gets better with having looser outage threshold. The energy efficiency behavior with respect to network density, SIR threshold and the number of retransmissions was also studied. In addition to the above results, we also show that the approximation used in our previous work for the average number of retransmissions is very precise and has very low error in the range.
\bibliographystyle{IEEEtran}
\bibliography{IEEEabrv,refsURC}

\begin{thebibliography}{10}
\providecommand{\url}[1]{#1}
\csname url@samestyle\endcsname
\providecommand{\newblock}{\relax}
\providecommand{\bibinfo}[2]{#2}
\providecommand{\BIBentrySTDinterwordspacing}{\spaceskip=0pt\relax}
\providecommand{\BIBentryALTinterwordstretchfactor}{4}
\providecommand{\BIBentryALTinterwordspacing}{\spaceskip=\fontdimen2\font plus
\BIBentryALTinterwordstretchfactor\fontdimen3\font minus
  \fontdimen4\font\relax}
\providecommand{\BIBforeignlanguage}[2]{{%
\expandafter\ifx\csname l@#1\endcsname\relax
\typeout{** WARNING: IEEEtran.bst: No hyphenation pattern has been}%
\typeout{** loaded for the language `#1'. Using the pattern for}%
\typeout{** the default language instead.}%
\else
\language=\csname l@#1\endcsname
\fi
#2}}
\providecommand{\BIBdecl}{\relax}
\BIBdecl

\bibitem{cerwall2015ericsson}
P.~Cerwall, P.~Jonsson, R.~M{\"o}ller, S.~B{\"a}vertoft, S.~Carson, and
  I.~Godor, ``Ericsson mobility report,'' \emph{On the Pulse of the Networked
  Society. Hg. v. Ericsson}, 2015.

\bibitem{manyika2015unlocking}
J.~Manyika, M.~Chui, P.~Bisson, J.~Woetzel, R.~Dobbs, J.~Bughin, and D.~Aharon,
  ``Unlocking the potential of the internet of things,'' \emph{McKinsey Global
  Institute}, 2015.

\bibitem{7004894}
C.~Perera, C.~H. Liu, S.~Jayawardena, and M.~Chen, ``A survey on internet of
  things from industrial market perspective,'' \emph{IEEE Access}, vol.~2, pp.
  1660--1679, 2014.

\bibitem{ali2015next}
A.~Ali, W.~Hamouda, and M.~Uysal, ``Next generation m2m cellular networks:
  challenges and practical considerations,'' \emph{IEEE Communications
  Magazine}, vol.~53, no.~9, pp. 18--24, 2015.

\bibitem{8476595}
P.~Popovski, K.~F. Trillingsgaard, O.~Simeone, and G.~Durisi, ``5g wireless
  network slicing for embb, urllc, and mmtc: A communication-theoretic view,''
  \emph{IEEE Access}, vol.~6, pp. 55\,765--55\,779, 2018.

\bibitem{dawy2017toward}
Z.~Dawy, W.~Saad, A.~Ghosh, J.~G. Andrews, and E.~Yaacoub, ``Toward massive
  machine type cellular communications,'' \emph{IEEE Wireless Communications},
  vol.~24, no.~1, pp. 120--128, 2017.

\bibitem{bockelmann2016massive}
C.~Bockelmann, N.~Pratas, H.~Nikopour, K.~Au, T.~Svensson, C.~Stefanovic,
  P.~Popovski, and A.~Dekorsy, ``Massive machine-type communications in 5g:
  Physical and mac-layer solutions,'' \emph{IEEE Communications Magazine},
  vol.~54, no.~9, pp. 59--65, 2016.

\bibitem{durisi2016toward}
G.~Durisi, T.~Koch, and P.~Popovski, ``Toward massive, ultrareliable, and
  low-latency wireless communication with short packets,'' \emph{Proceedings of
  the IEEE}, vol. 104, no.~9, pp. 1711--1726, 2016.

\bibitem{akyildiz2006next}
I.~F. Akyildiz, W.-Y. Lee, M.~C. Vuran, and S.~Mohanty, ``Next
  generation/dynamic spectrum access/cognitive radio wireless networks: A
  survey,'' \emph{Computer networks}, vol.~50, no.~13, pp. 2127--2159, 2006.

\bibitem{nardelli2016maximizing}
P.~H. Nardelli, M.~de~Castro~Tom{\'e}, H.~Alves, C.~H. de~Lima, and
  M.~Latva-aho, ``Maximizing the link throughput between smart meters and
  aggregators as secondary users under power and outage constraints,'' \emph{Ad
  Hoc Networks}, vol.~41, pp. 57--68, 2016.

\bibitem{tome2016joint}
M.~C. Tom{\'e}, P.~H. Nardelli, H.~Alves, and M.~Latva-aho, ``Joint
  sampling-communication strategies for smart-meters to aggregator link as
  secondary users,'' in \emph{Energy Conference (ENERGYCON), 2016 IEEE
  International}.\hskip 1em plus 0.5em minus 0.4em\relax IEEE, 2016, pp. 1--6.

\bibitem{ramezanipour2018increasing}
I.~Ramezanipour, P.~H. Nardelli, H.~Alves, and A.~Pouttu, ``Increasing the
  throughput of an unlicensed wireless network through retransmissions,'' in
  \emph{2018 IEEE 87th Vehicular Technology Conference (VTC Spring)}, 2018, pp.
  1--5.

\bibitem{nardelli2012optimal}
P.~H. Nardelli, M.~Kaynia, P.~Cardieri, and M.~Latva-aho, ``Optimal
  transmission capacity of ad hoc networks with packet retransmissions,''
  \emph{IEEE Transactions on Wireless Communications}, vol.~11, no.~8, pp.
  2760--2766, 2012.

\bibitem{nardelli2014throughput}
P.~H. Nardelli, M.~Kountouris, P.~Cardieri, and M.~Latva-Aho, ``Throughput
  optimization in wireless networks under stability and packet loss
  constraints,'' \emph{IEEE Transactions on Mobile Computing}, vol.~13, no.~8,
  pp. 1883--1895, 2014.

\bibitem{iran}
I.~Ramezanipour, P.~H. Nardelli, H.~Alves, and A.~Pouttu, ``Energy efficiency
  of an unlicensed wireless network in the presence of retransmissions,'' in
  \emph{2018 IEEE 87th Vehicular Technology Conference (VTC Spring)}, 2018.

\bibitem{osseiran2014scenarios}
A.~Osseiran, F.~Boccardi, V.~Braun, K.~Kusume, P.~Marsch, M.~Maternia,
  O.~Queseth, M.~Schellmann, H.~Schotten, H.~Taoka \emph{et~al.}, ``Scenarios
  for 5g mobile and wireless communications: the vision of the metis project,''
  \emph{IEEE Communications Magazine}, vol.~52, no.~5, pp. 26--35, 2014.

\bibitem{popovski20185g}
P.~Popovski, K.~F. Trillingsgaard, O.~Simeone, and G.~Durisi, ``5g wireless
  network slicing for embb, urllc, and mmtc: A communication-theoretic view,''
  \emph{arXiv preprint arXiv:1804.05057}, 2018.

\bibitem{akyildiz2002wireless}
I.~F. Akyildiz, W.~Su, Y.~Sankarasubramaniam, and E.~Cayirci, ``Wireless sensor
  networks: a survey,'' \emph{Computer networks}, vol.~38, no.~4, pp. 393--422,
  2002.

\bibitem{de2011energy}
G.~G. de~Oliveira~Brante, M.~T. Kakitani, and R.~D. Souza, ``Energy efficiency
  analysis of some cooperative and non-cooperative transmission schemes in
  wireless sensor networks,'' \emph{IEEE Transactions on Communications},
  vol.~59, no.~10, pp. 2671--2677, 2011.

\bibitem{vardhan2000wireless}
S.~Vardhan, M.~Wilczynski, G.~Portie, and W.~J. Kaiser, ``Wireless integrated
  network sensors (wins): distributed in situ sensing for mission and flight
  systems,'' in \emph{Aerospace Conference Proceedings, 2000 IEEE},
  vol.~7.\hskip 1em plus 0.5em minus 0.4em\relax IEEE, 2000, pp. 459--463.

\bibitem{hasan2011green}
Z.~Hasan, H.~Boostanimehr, and V.~K. Bhargava, ``Green cellular networks: A
  survey, some research issues and challenges,'' \emph{IEEE Communications
  surveys \& tutorials}, vol.~13, no.~4, pp. 524--540, 2011.

\bibitem{wang2010energy}
S.~Wang and J.~Nie, ``Energy efficiency optimization of cooperative
  communication in wireless sensor networks,'' \emph{EURASIP Journal on
  Wireless Communications and Networking}, vol. 2010, p.~3, 2010.

\bibitem{sadek2009energy}
A.~K. Sadek, W.~Yu, and K.~Liu, ``On the energy efficiency of cooperative
  communications in wireless sensor networks,'' \emph{ACM Transactions on
  Sensor Networks (TOSN)}, vol.~6, no.~1, p.~5, 2009.

\bibitem{zhihui2014eefa}
G.~Zhihui, L.~Anzhong, and L.~Taoshen, ``Eefa: Energy efciency frame
  aggregation scheduling algorithm for ieee 802.11 n wireless network,''
  \emph{China Communications}, vol.~11, no.~3, pp. 19--26, 2014.

\bibitem{ma2007battery}
C.~Ma and Y.~Yang, ``A battery aware scheme for energy efficient coverage and
  routing in wireless mesh networks,'' in \emph{Global Telecommunications
  Conference, 2007. GLOBECOM'07. IEEE}.\hskip 1em plus 0.5em minus 0.4em\relax
  IEEE, 2007, pp. 1113--1117.

\bibitem{jayashree2004battery}
S.~Jayashree, B.~Manoj, and C.~S.~R. Murthy, ``A battery aware medium access
  control (bamac) protocol for ad hoc wireless networks,'' in \emph{Personal,
  Indoor and Mobile Radio Communications, 2004. PIMRC 2004. 15th IEEE
  International Symposium on}, vol.~2.\hskip 1em plus 0.5em minus 0.4em\relax
  IEEE, 2004, pp. 995--999.

\bibitem{kitahara2008data}
T.~Kitahara and H.~Nakamura, ``A data transmission control to maximize
  discharge capacity of battery,'' in \emph{Wireless Communications and
  Networking Conference, 2008. WCNC 2008. IEEE}.\hskip 1em plus 0.5em minus
  0.4em\relax IEEE, 2008, pp. 2951--2956.

\bibitem{tu2011energy}
C.-Y. Tu, C.-Y. Ho, and C.-Y. Huang, ``Energy-efficient algorithms and
  evaluations for massive access management in cellular based machine to
  machine communications,'' in \emph{Vehicular Technology Conference (VTC
  Fall), 2011 IEEE}.\hskip 1em plus 0.5em minus 0.4em\relax IEEE, 2011, pp.
  1--5.

\bibitem{joint}
J.~Wildman, P.~H.~J. Nardelli, M.~Latva-aho, and S.~Weber, ``On the joint
  impact of beamwidth and orientation error on throughput in directional
  wireless poisson networks,'' \emph{IEEE Transactions on Wireless
  Communications}, vol.~13, no.~12, pp. 7072--7085, 2014.

\bibitem{haenggi2012stochastic}
M.~Haenggi, \emph{Stochastic geometry for wireless networks}.\hskip 1em plus
  0.5em minus 0.4em\relax Cambridge University Press, 2012.

\bibitem{iranj}
I.~Ramezanipour, P.~Nouri, H.~Alves, P.~H. Nardelli, R.~D. Souza, and
  A.~Pouttu, ``Finite blocklength communications in smart grids for dynamic
  spectrum access and locally licensed scenarios,'' \emph{IEEE Sensors
  Journal}, vol.~18, no.~13, pp. 5610--5621, 2018.

\bibitem{weber2010overview}
S.~Weber, J.~G. Andrews, and N.~Jindal, ``An overview of the transmission
  capacity of wireless networks,'' \emph{IEEE Transactions on Communications},
  vol.~58, no.~12, pp. 3593--3604, 2010.

\bibitem{nardelli2015throughput}
P.~H. Nardelli, C.~H. de~Lima, H.~Alves, P.~Cardieri, and M.~Latva-aho,
  ``Throughput analysis of cognitive wireless networks with poisson distributed
  nodes based on location information,'' \emph{Ad Hoc Networks}, vol.~33, pp.
  1--15, 2015.

\bibitem{baccelli2011interference}
F.~Baccelli, A.~El~Gamal, and N.~David, ``Interference networks with
  point-to-point codes,'' \emph{IEEE Transactions on Information Theory},
  vol.~57, no.~5, pp. 2582--2596, 2011.

\bibitem{p2}
P.~H. Nardelli, M.~Kaynia, P.~Cardieri, and M.~Latva-aho, ``Optimal
  transmission capacity of ad hoc networks with packet retransmissions,''
  \emph{IEEE Transactions on Wireless Communications}, vol.~11, no.~8, pp.
  2760--2766, 2012.

\bibitem{nardelli2009multi}
P.~H.~J. Nardelli, G.~T.~F. de~Abreu, and P.~Cardieri, ``Multi-hop aggregate
  information efficiency in wireless ad hoc networks,'' in \emph{IEEE
  International Conference on Communications, 2009. ICC'09.}\hskip 1em plus
  0.5em minus 0.4em\relax IEEE, 2009, pp. 1--6.

\bibitem{haenggi2013local}
M.~Haenggi, ``The local delay in poisson networks,'' \emph{IEEE Transactions on
  Information Theory}, vol.~59, no.~3, pp. 1788--1802, 2013.

\bibitem{cui2005energy}
S.~Cui, A.~J. Goldsmith, and A.~Bahai, ``Energy-constrained modulation
  optimization,'' \emph{IEEE transactions on wireless communications}, vol.~4,
  no.~5, pp. 2349--2360, 2005.

\bibitem{i1}
I.~Ramezanipour, H.~Alves, P.~H.~J. Nardelli, and A.~Pouttu, ``Energy
  efficiency of an unlicensed wireless network in the presence of
  retransmissions,'' in \emph{2018 IEEE 87th Vehicular Technology Conference
  (VTC Spring)}, June 2018, pp. 1--5.

\bibitem{alves2014outage}
H.~Alves, R.~D. Souza, and G.~Fraidenraich, ``Outage, throughput and energy
  efficiency analysis of some half and full duplex cooperative relaying
  schemes,'' \emph{Transactions on Emerging Telecommunications Technologies},
  vol.~25, no.~11, pp. 1114--1125, 2014.

\end{thebibliography}


@inproceedings{ramezanipour2018increasing,
  title={Increasing the throughput of an unlicensed wireless network through retransmissions},
  author={Ramezanipour, Iran and Nardelli, Pedro HJ and Alves, Hirley and Pouttu, Ari},
  booktitle={2018 IEEE 87th Vehicular Technology Conference (VTC Spring)},
  pages={1--5},
  year={2018},
  
}

@inproceedings{iran,
	title={Energy Efficiency of an Unlicensed Wireless Network in the Presence of Retransmissions},
	author={Ramezanipour, Iran and Nardelli, Pedro HJ and Alves, Hirley and Pouttu, Ari},
	booktitle={2018 IEEE 87th Vehicular Technology Conference (VTC Spring)},
	year={2018},
	
}

@article{nardelli2016maximizing,
	title={Maximizing the link throughput between smart meters and aggregators as secondary users under power and outage constraints},
	author={Nardelli, Pedro HJ and de Castro Tom{\'e}, Mauricio and Alves, Hirley and de Lima, Carlos HM and Latva-aho, Matti},
	journal={Ad Hoc Networks},
	volume={41},
	pages={57--68},
	year={2016},
	publisher={Elsevier}
}

@article{joint,
	title={On the joint impact of beamwidth and orientation error on throughput in directional wireless Poisson networks},
	author={Wildman, Jeffrey and Nardelli, Pedro Henrique Juliano and Latva-aho, Matti and Weber, Steven},
	journal={IEEE Transactions on Wireless Communications},
	volume={13},
	number={12},
	pages={7072--7085},
	year={2014},
	publisher={IEEE}
}

@book{haenggi2012stochastic,
	title={Stochastic geometry for wireless networks},
	author={Haenggi, Martin},
	year={2012},
	publisher={Cambridge University Press}
}

@article{iranj,
	title={Finite Blocklength Communications in Smart Grids for Dynamic Spectrum Access and Locally Licensed Scenarios},
	author={Ramezanipour, Iran and Nouri, Parisa and Alves, Hirley and Nardelli, Pedro HJ and Souza, Richard Demo and Pouttu, Ari},
	journal={IEEE Sensors Journal},
	volume={18},
	number={13},
	pages={5610--5621},
	year={2018},
	publisher={IEEE}
}

@article{weber2010overview,
	title={An overview of the transmission capacity of wireless networks},
	author={Weber, Steven and Andrews, Jeffrey G and Jindal, Nihar},
	journal={IEEE Transactions on Communications},
	volume={58},
	number={12},
	pages={3593--3604},
	year={2010},
	publisher={IEEE}
}

@article{nardelli2015throughput,
	title={Throughput analysis of cognitive wireless networks with Poisson distributed nodes based on location information},
	author={Nardelli, Pedro HJ and de Lima, Carlos HM and Alves, Hirley and Cardieri, Paulo and Latva-aho, Matti},
	journal={Ad Hoc Networks},
	volume={33},
	pages={1--15},
	year={2015},
	publisher={Elsevier}
}

@article{baccelli2011interference,
	title={Interference networks with point-to-point codes},
	author={Baccelli, Francois and El Gamal, Abbas and David, NC},
	journal={IEEE Transactions on Information Theory},
	volume={57},
	number={5},
	pages={2582--2596},
	year={2011},
	publisher={IEEE}
}

@article{p2,
	title={Optimal transmission capacity of ad hoc networks with packet retransmissions},
	author={Nardelli, Pedro HJ and Kaynia, Mariam and Cardieri, Paulo and Latva-aho, Matti},
	journal={IEEE Transactions on Wireless Communications},
	volume={11},
	number={8},
	pages={2760--2766},
	year={2012},
	publisher={IEEE}
}

@inproceedings{nardelli2009multi,
	title={Multi-hop aggregate information efficiency in wireless ad hoc networks},
	author={Nardelli, Pedro Henrique Juliano and de Abreu, Giuseppe Thadeu Freitas and Cardieri, Paulo},
	booktitle={IEEE International Conference on Communications, 2009. ICC'09.},
	pages={1--6},
	year={2009},
	organization={IEEE}
}

@INPROCEEDINGS{i1, 
	author={I. Ramezanipour and H. Alves and P. H. J. Nardelli and A. Pouttu}, 
	booktitle={2018 IEEE 87th Vehicular Technology Conference (VTC Spring)}, 
	title={Energy Efficiency of an Unlicensed Wireless Network in the Presence of Retransmissions}, 
	year={2018}, 
	volume={}, 
	number={}, 
	pages={1-5}, 
	keywords={decoding;energy conservation;optimisation;power consumption;radiofrequency interference;stochastic processes;telecommunication network reliability;telecommunication power management;wireless sensor networks;outage threshold;power consumption;Poisson point process;uplink channel;unlicensed users;sensor nodes;wireless sensor network;transmitted message;unlicensed wireless network;energy efficiency;network density;Throughput;Wireless sensor networks;Interference;Receivers;Wireless networks;Power demand;Numerical models}, 
	doi={10.1109/VTCSpring.2018.8417869}, 
	ISSN={2577-2465}, 
	month={June},}

@article{de2011energy,
	title={Energy efficiency analysis of some cooperative and non-cooperative transmission schemes in wireless sensor networks},
	author={de Oliveira Brante, Glauber Gomes and Kakitani, Marcos Tomio and Souza, Richard Demo},
	journal={IEEE Transactions on Communications},
	volume={59},
	number={10},
	pages={2671--2677},
	year={2011},
	publisher={IEEE}
}

@article{alves2014outage,
	title={Outage, throughput and energy efficiency analysis of some half and full duplex cooperative relaying schemes},
	author={Alves, Hirley and Souza, Richard Demo and Fraidenraich, Gustavo},
	journal={Transactions on Emerging Telecommunications Technologies},
	volume={25},
	number={11},
	pages={1114--1125},
	year={2014},
	publisher={Wiley Online Library}
}

@article{haenggi2013local,
	title={The local delay in Poisson networks},
	author={Haenggi, Martin},
	journal={IEEE Transactions on Information Theory},
	volume={59},
	number={3},
	pages={1788--1802},
	year={2013},
	publisher={IEEE}
}

@article{cerwall2015ericsson,
	title={Ericsson mobility report},
	author={Cerwall, Patrik and Jonsson, P and M{\"o}ller, R and B{\"a}vertoft, S and Carson, S and Godor, I},
	journal={On the Pulse of the Networked Society. Hg. v. Ericsson},
	year={2015}
}

@article{manyika2015unlocking,
	title={Unlocking the Potential of the Internet of Things},
	author={Manyika, James and Chui, Michael and Bisson, Peter and Woetzel, Jonathan and Dobbs, Richard and Bughin, Jacques and Aharon, Dan},
	journal={McKinsey Global Institute},
	year={2015}
}

@ARTICLE{7004894, 
	author={C. Perera and C. H. Liu and S. Jayawardena and M. Chen}, 
	journal={IEEE Access}, 
	title={A Survey on Internet of Things From Industrial Market Perspective}, 
	year={2014}, 
	volume={2}, 
	number={}, 
	pages={1660-1679}, 
	keywords={industrial economics;Internet of Things;product development;production engineering computing;context-aware product development;ubiquitous computing;industrial market;IoT;Internet of Things;Context awareness;Internet of things;Market research;Information networks;Internet;Globalization;Interconnections;Product development;Futures research;Internet of things;industry solutions;contextawareness;product review;IoT marketplace;Internet of Things;industry solutions;context-awareness;product review;IoT marketplace}, 
	doi={10.1109/ACCESS.2015.2389854}, 
	ISSN={2169-3536}, 
	month={},}

@article{ali2015next,
	title={Next generation M2M cellular networks: challenges and practical considerations},
	author={Ali, Abdelmohsen and Hamouda, Walaa and Uysal, Murat},
	journal={IEEE Communications Magazine},
	volume={53},
	number={9},
	pages={18--24},
	year={2015},
	publisher={IEEE}
}

@article{bockelmann2016massive,
	title={Massive machine-type communications in 5G: Physical and MAC-layer solutions},
	author={Bockelmann, Carsten and Pratas, Nuno and Nikopour, Hosein and Au, Kelvin and Svensson, Tommy and Stefanovic, Cedomir and Popovski, Petar and Dekorsy, Armin},
	journal={IEEE Communications Magazine},
	volume={54},
	number={9},
	pages={59--65},
	year={2016},
	publisher={IEEE}
}

@article{durisi2016toward,
	title={Toward massive, ultrareliable, and low-latency wireless communication with short packets},
	author={Durisi, Giuseppe and Koch, Tobias and Popovski, Petar},
	journal={Proceedings of the IEEE},
	volume={104},
	number={9},
	pages={1711--1726},
	year={2016},
	publisher={IEEE}
}

@article{akyildiz2006next,
	title={NeXt generation/dynamic spectrum access/cognitive radio wireless networks: A survey},
	author={Akyildiz, Ian F and Lee, Won-Yeol and Vuran, Mehmet C and Mohanty, Shantidev},
	journal={Computer networks},
	volume={50},
	number={13},
	pages={2127--2159},
	year={2006},
	publisher={Elsevier}
}

@inproceedings{tome2016joint,
	title={Joint sampling-communication strategies for smart-meters to aggregator link as secondary users},
	author={Tom{\'e}, Mauricio C and Nardelli, Pedro HJ and Alves, Hirley and Latva-aho, Matti},
	booktitle={Energy Conference (ENERGYCON), 2016 IEEE International},
	pages={1--6},
	year={2016},
	organization={IEEE}
}

@article{nardelli2012optimal,
	title={Optimal transmission capacity of ad hoc networks with packet retransmissions},
	author={Nardelli, Pedro HJ and Kaynia, Mariam and Cardieri, Paulo and Latva-aho, Matti},
	journal={IEEE Transactions on Wireless Communications},
	volume={11},
	number={8},
	pages={2760--2766},
	year={2012},
	publisher={IEEE}
}

@article{nardelli2014throughput,
	title={Throughput optimization in wireless networks under stability and packet loss constraints},
	author={Nardelli, Pedro HJ and Kountouris, Marios and Cardieri, Paulo and Latva-Aho, Matti},
	journal={IEEE Transactions on Mobile Computing},
	volume={13},
	number={8},
	pages={1883--1895},
	year={2014},
	publisher={IEEE}
}

@article{osseiran2014scenarios,
	title={Scenarios for 5G mobile and wireless communications: the vision of the METIS project},
	author={Osseiran, Afif and Boccardi, Federico and Braun, Volker and Kusume, Katsutoshi and Marsch, Patrick and Maternia, Michal and Queseth, Olav and Schellmann, Malte and Schotten, Hans and Taoka, Hidekazu and others},
	journal={IEEE Communications Magazine},
	volume={52},
	number={5},
	pages={26--35},
	year={2014},
	publisher={IEEE}
}

@article{popovski20185g,
	title={5G Wireless Network Slicing for eMBB, URLLC, and mMTC: A Communication-Theoretic View},
	author={Popovski, Petar and Trillingsgaard, Kasper F and Simeone, Osvaldo and Durisi, Giuseppe},
	journal={arXiv preprint arXiv:1804.05057},
	year={2018}
}

@article{akyildiz2002wireless,
	title={Wireless sensor networks: a survey},
	author={Akyildiz, Ian F and Su, Weilian and Sankarasubramaniam, Yogesh and Cayirci, Erdal},
	journal={Computer networks},
	volume={38},
	number={4},
	pages={393--422},
	year={2002},
	publisher={Elsevier}
}

@article{de2011energy,
	title={Energy efficiency analysis of some cooperative and non-cooperative transmission schemes in wireless sensor networks},
	author={de Oliveira Brante, Glauber Gomes and Kakitani, Marcos Tomio and Souza, Richard Demo},
	journal={IEEE Transactions on Communications},
	volume={59},
	number={10},
	pages={2671--2677},
	year={2011},
	publisher={IEEE}
}

@inproceedings{vardhan2000wireless,
	title={Wireless integrated network sensors (WINS): distributed in situ sensing for mission and flight systems},
	author={Vardhan, Sandeep and Wilczynski, Matt and Portie, GJ and Kaiser, William J},
	booktitle={Aerospace Conference Proceedings, 2000 IEEE},
	volume={7},
	pages={459--463},
	year={2000},
	organization={IEEE}
}

@article{hasan2011green,
	title={Green cellular networks: A survey, some research issues and challenges},
	author={Hasan, Ziaul and Boostanimehr, Hamidreza and Bhargava, Vijay K},
	journal={IEEE Communications surveys \& tutorials},
	volume={13},
	number={4},
	pages={524--540},
	year={2011},
	publisher={IEEE}
}

@article{wang2010energy,
	title={Energy efficiency optimization of cooperative communication in wireless sensor networks},
	author={Wang, Shaoqing and Nie, Jingnan},
	journal={EURASIP Journal on Wireless Communications and Networking},
	volume={2010},
	pages={3},
	year={2010},
	publisher={Hindawi Publishing Corp.}
}

@article{sadek2009energy,
	title={On the energy efficiency of cooperative communications in wireless sensor networks},
	author={Sadek, Ahmed K and Yu, Wei and Liu, KJ},
	journal={ACM Transactions on Sensor Networks (TOSN)},
	volume={6},
	number={1},
	pages={5},
	year={2009},
	publisher={ACM}
}

@inproceedings{ma2007battery,
	title={A battery aware scheme for energy efficient coverage and routing in wireless mesh networks},
	author={Ma, Chi and Yang, Yuanyuan},
	booktitle={Global Telecommunications Conference, 2007. GLOBECOM'07. IEEE},
	pages={1113--1117},
	year={2007},
	organization={IEEE}
}

@inproceedings{jayashree2004battery,
	title={A battery aware medium access control (BAMAC) protocol for ad hoc wireless networks},
	author={Jayashree, Subramanian and Manoj, BS and Murthy, C Siva Ram},
	booktitle={Personal, Indoor and Mobile Radio Communications, 2004. PIMRC 2004. 15th IEEE International Symposium on},
	volume={2},
	pages={995--999},
	year={2004},
	organization={IEEE}
}

@inproceedings{kitahara2008data,
	title={A data transmission control to maximize discharge capacity of battery},
	author={Kitahara, Takeshi and Nakamura, Hajime},
	booktitle={Wireless Communications and Networking Conference, 2008. WCNC 2008. IEEE},
	pages={2951--2956},
	year={2008},
	organization={IEEE}
}

@inproceedings{tu2011energy,
	title={Energy-efficient algorithms and evaluations for massive access management in cellular based machine to machine communications},
	author={Tu, Chih-Yuan and Ho, Chieh-Yuan and Huang, Ching-Yao},
	booktitle={Vehicular Technology Conference (VTC Fall), 2011 IEEE},
	pages={1--5},
	year={2011},
	organization={IEEE}
}

@article{cui2005energy,
	title={Energy-constrained modulation optimization},
	author={Cui, Shuguang and Goldsmith, Andrea J and Bahai, Ahmad},
	journal={IEEE transactions on wireless communications},
	volume={4},
	number={5},
	pages={2349--2360},
	year={2005},
	publisher={IEEE}
}

@ARTICLE{8476595, 
	author={P. Popovski and K. F. Trillingsgaard and O. Simeone and G. Durisi}, 
	journal={IEEE Access}, 
	title={5G Wireless Network Slicing for eMBB, URLLC, and mMTC: A Communication-Theoretic View}, 
	year={2018}, 
	volume={6}, 
	number={}, 
	pages={55765-55779}, 
	keywords={5G mobile communication;multi-access systems;radio access networks;resource allocation;telecommunication network reliability;5G wireless network slicing;eMBB;communication-theoretic view;generic services;vastly heterogeneous requirements;massive machine-type communications;service heterogeneity;radio access network;orthogonal resource allocation;nonorthogonal sharing;RAN resources;uplink communications;URLLC devices;common base station;heterogeneous nonorthogonal multiple access;H-NOMA;conventional NOMA techniques;homogeneous requirements;standard multiple access channel;communication-theoretic model;reliability diversity;nonorthogonal RAN slicing;orthogonal slicing;ultra-reliable low-latency communications;reliability requirements;NOMA;Reliability;Wireless communication;Resource management;Radio spectrum management;Time-frequency analysis;5G mobile communication;5G mobile communication;machine-to-machine communications;multiaccess communication;NOMA;wireless communication}, 
	doi={10.1109/ACCESS.2018.2872781}, 
	ISSN={2169-3536}, 
	month={},}

@article{dawy2017toward,
	title={Toward massive machine type cellular communications},
	author={Dawy, Zaher and Saad, Walid and Ghosh, Arunabha and Andrews, Jeffrey G and Yaacoub, Elias},
	journal={IEEE Wireless Communications},
	volume={24},
	number={1},
	pages={120--128},
	year={2017},
	publisher={IEEE}
}

@article{zhihui2014eefa,
	title={EEFA: Energy efciency frame aggregation scheduling algorithm for IEEE 802.11 n wireless network},
	author={Zhihui, Ge and Anzhong, Liang and Taoshen, Li},
	journal={China Communications},
	volume={11},
	number={3},
	pages={19--26},
	year={2014},
	publisher={IEEE}
}
\end{document}